\documentclass[aps,prl,reprint,superscriptaddress,amsmath,amssymb,floatfix,showpacs]{revtex4-1}
\usepackage{mathtools}
\usepackage{amsmath}
\usepackage{graphicx}
\usepackage{esint}
\usepackage{bm}

\usepackage{hyperref}

\makeatletter
\usepackage{color}
\usepackage{epstopdf}
\usepackage{slashed}
\usepackage{bm}
\usepackage{braket}
\usepackage{soul}

\makeatother
\begin{document}

\title{Quantum-critical scaling at the Bose-glass transition of the 3d diluted Heisenberg antiferromagnet in a field}

\author{Yuchen Fan}
\affiliation{Department of Physics and Beijing Key Laboratory of Opto-electronic Functional Materials and Micro-nano Devices, Renmin University of China, Beijing 100872, China}
\affiliation{Beijing National Laboratory for Condensed Matter Physics and Institute of Physics,
Chinese Academy of Sciences, Beijing, 100190, China}

\author{Rong Yu}
\email{rong.yu@ruc.edu.cn}
\affiliation{Department of Physics and Beijing Key Laboratory of Opto-electronic Functional Materials and Micro-nano Devices, Renmin University of China, Beijing 100872, China}

\author{Tommaso Roscilde}
\email{tommaso.roscilde@ens-lyon.fr}
\affiliation{Univ Lyon, Ens de Lyon, CNRS, Laboratoire de Physique, F-69342 Lyon, France}

\begin{abstract}
The nature of the superfluid-to-Bose-glass (SF-BG) quantum phase transition, occurring in systems of interacting bosons immersed in a disordered environment, remains elusive. One fundamental open question is whether or not the transition obeys conventional scaling at quantum critical points (QCPs): this scaling would lock the value of the crossover exponent $\phi$ -- dictating the vanishing of the superfluid critical temperature upon approaching the QCP -- to the value of quantum critical exponents for the ground-state transition. Yet such a relation between exponents has been called into question by several numerical as well as experimental results on the SF-BG transition. Here we revisit this issue in the case of the $S=1/2$ Heisenberg antiferromagnet on a site-diluted cubic lattice, which lends itself to efficient quantum Monte Carlo simulations. Our results show that the model exhibits a percolation transition in zero applied field, with the correlation length exponent $\nu = 0.87(8)$ 
and $\phi = 1.1(1)$ consistent with 3d percolation. When applying a sufficiently strong magnetic field, the dilution-induced transition decouples from geometric percolation, and it becomes a SF-BG transition; nonetheless, the $\nu$ and $\phi$ exponents maintain values consistent with those of the percolation transition. These results contradict the conventional scaling, which predicts $\phi\geqslant 2$; and they suggest a close connection between the SF-BG transition and percolation.
\end{abstract}
\maketitle

\textit{Introduction.}  Quantum localization in strongly correlated systems is a central topic of current research in condensed matter and statistical physics \cite{GiamarchiS1987, Fisheretal1989, Prokofevbook, Abrahamsbook, Dobrosavljevicbook,Vojta2019}. The phenomenon of localization in interacting systems has wide-ranging consequences on the equilibrium physics of the system (which is the focus of this work) as well as on its non-equilibrium dynamics, with the possible breakdown of thermalization \cite{AletL2018,Abaninetal2019}.  In this work we shall specifically focus on the localization transition in bosonic systems, which is relevant to a wide variety of physical platforms, ranging from $^4$He in porous media \cite{Crowelletal1997}, to ultracold atoms in disordered optical potentials \cite{DErricoetal2014} to disordered superconductors \cite{Sacepeetal2011}. The central question in this domain concerns the nature of the ground-state quantum phase transition (QPT)\cite{GiamarchiS1987, Fisheretal1989} from the long-range ordered, superfluid (SF) phase  -- for weak disorder and/or sufficiently strong interactions; to the localized, Bose-glass (BG) phase -- for strong disorder and/or weak interactions.
The nature of the QPT remains a subject of debate, as it appears to defy the conventional scenario for critical scaling. The seminal work of Ref.~\cite{Fisheretal1989} postulated that the finite-temperature free energy in the vicinity of the QPT obeys a conventional scaling form, admitting the temperature and the control parameter of the QPT as the unique instability directions for the fixed point governing the quantum critical behavior.
This has the result that, on the SF side of the transition, long-range order persists in 3d up to a critical temperature $T_c$ which vanishes upon approaching the quantum-critical point (QCP) as $T_c \sim |g-g_c|^{\phi}$ (with $g$ the control parameter of the QPT), with a crossover exponent $\phi = \nu z$.  Here $\nu$ is the correlation-length exponent at the QCP, and $z$ the dynamical critical exponent. The finite compressibility of the BG phase leads to the prediction that $z=d$ \cite{Fisheretal1989} with $d$ the number of dimensions. Combined with the Harris criterion for criticality in disordered systems \cite{Vojta2019}, mandating that $\nu \geq 2/d$, this leads to the conclusion that $\phi \geq 2$.

The above picture has been severely cast into doubt by the first quantitative tests of the scaling theory for the SF-BG transition, which have been mostly performed on its magnetic incarnation \cite{ZheludevR2013}. A broad family of gapped magnetic systems exhibits a magnetic analog of the commensurate-incommensurate transition of strongly interacting bosons \cite{Sachdevbook} when subject to a strong magnetic field \cite{Giamarchietal2008,Zapfetal2014}; and this transition takes the nature of a SF-BG transition when disorder is introduced in the system  \cite{ZheludevR2013}. The magnetic SF-BG transition has been investigated in a variety of doped compounds \cite{Manakaetal2008,Hongetal2010,Yamadaetal2011,Huvonenetal2012,Yuetal2012} and theoretical models thereof \cite{Yuetal2012,Yuetal2012-b}, the most extensively studied case being the one of NiCl$_{2(1-x)}$Br$_{2x}\cdot4$SC(NH$_2$)$_2$ (Br-DTN) \cite{Yuetal2010,Yuetal2012,Yuetal2012-b}. All these studies have concluded that  $\phi<2$, with the most in-depth investigations giving $\phi\approx 1.1$, at least in the range of $T_c$ that were accessible to both experiments and numerics. This is in open contradiction with the $\phi \geq 2$ prediction, casting one or more hypothesis behind it into doubt. The Harris criterion $\nu \geq 2/3$ \cite{Vojta2019} is confirmed by all the studies of the 3d SF-BG transition \cite{HitchcockS2006,Yuetal2012-b,Yaoetal2014}, giving $\nu \approx 0.7-0.9$. The prediction $z=d$, while called into question by several studies in $d=2$ \cite{MeierW2012,NgS2015}, appears to be confirmed by all the available studies of the 3d SF-BG transition \cite{HitchcockS2006,Yuetal2012-b,Yaoetal2014}. Therefore a possible conclusion is that the SF-BG transition does not obey conventional scaling, and several generalized two-argument scaling Ans\"atze have been put forward \cite{Yuetal2012-b}.

Meanwhile, a large-scale numerical study \cite{Yaoetal2014} found that $\phi \approx 2.7$ for the disordered link-current model (a dual model to the disordered Bose-Hubbard model) and for disordered hardcore bosons at zero average chemical potential. In particular Ref.~\cite{Yaoetal2014} puts forward a density criterion $|n(T)/n_c-1| \ll 1$ identifying the putative temperature range to observe the correct $\phi$ exponent, where $n(T)$ is the finite-$T$ boson density and $n_c$ its value at the QCP; and arguing that the  observation of $\phi \approx 1.1$ is the result of an analysis of $T_c$ values in a temperature range not complying with this criterion. Nonetheless this argument is contradicted by the fact that existing data consistent with $\phi\approx 1.1$ \cite{Yuetal2012} are also complying with the density criterion \cite{Yuetal2014}.

Here we revisit the $\phi$ exponent problem at the 3d SF-BG transition by looking at the $S=1/2$ Heisenberg antiferromagnet on a site-diluted cubic lattice and in a uniform magnetic field, which lends itself to very efficient quantum Monte Carlo (QMC) simulations of the transition on vey big lattices and at very low temperatures. Long-range magnetic order can be destroyed in the ground state by a sufficiently large dilution $x$ (density of empty sites), and in zero applied field this transition is found to have a percolative nature, exhibiting a crossover exponent  $\phi=1.1(1)$ consistent with the one for 3d percolation, $\phi_p = 1.12(2)$ \cite{percolationbook,Coniglio1981,Khairnaretal2021}. The transition becomes a SF-BG one in a finite field, but the crossover exponent is found to maintain a value consistent with the percolation one. Our data reverse the conclusions of Ref.~\cite{Yaoetal2014}, in that an exponent $\phi\gtrsim 2$ can only fit our $T_c$ data in an intermediate temperature range, but it cannot account for the lowest-temperature regime we access. Our results suggest the existence of a strong link between the Bose-glass transition and percolation, but in a way which is very strongly affected by quantum fluctuations at zero temperature \cite{Yuetal2010}.


\begin{figure}
	\includegraphics[width=0.7\columnwidth]{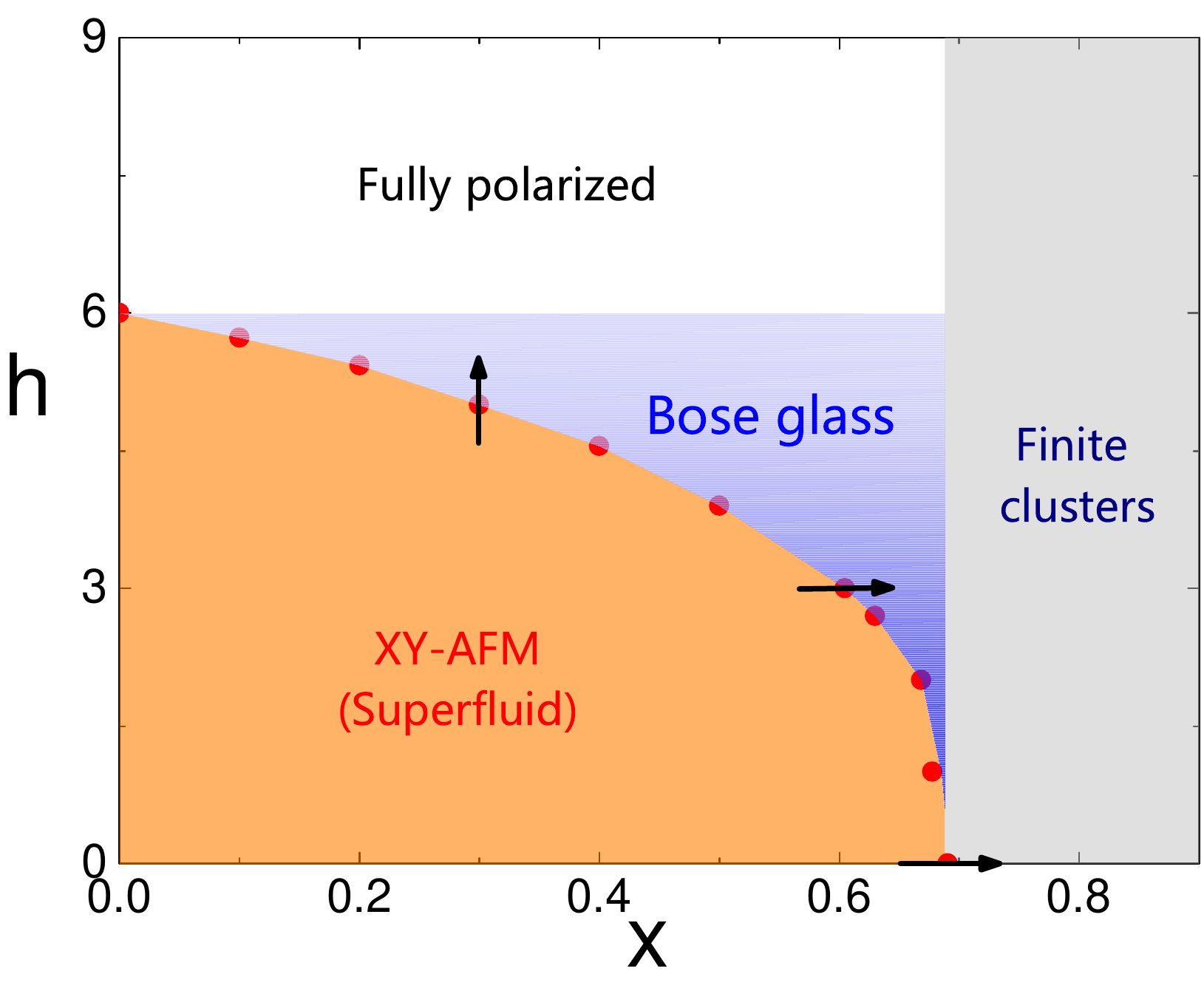} 
	\caption{Ground state phase diagram of the $S=1/2$ 3D site-diluted AFM Heisenberg model in the $x$-$h$ plane. The dots give the QMC estimates of the order-disorder transition at $t=10^{-3}$.  Arrows illustrate the three different transitions analyzed in details in this work.}
	\label{fig:1}
\end{figure}

\textit{Model and ground-state phase diagram.} We study the $S=1/2$ Heisenberg antiferromagnet defined on a site-diluted cubic lattice,  whose Hamiltonian reads
\begin{equation}\label{Eq:Ham}
\mathcal{H}=J \sum_{\langle i,j \rangle}{\epsilon_i \epsilon_j {\bm S}_i\cdot {\bm S}_j}- H\sum_i{\epsilon_i S_{i}^{z}};
\end{equation}
here ${\bm S}_i$ is an $S = 1/2$ quantum spin operator on each site $i$; $H$ is the applied magnetic field; and
$\epsilon_i$ takes values $1$ and $0$ randomly, with probabilities $1-x$ and $x$ respectively -- where $x$ is the dilution concentration.   
This model is also equivalent to hardcore bosons with hopping $J/2$, nearest neighbor repulsion $J$, and a uniform chemical potential $H$, defined on the same lattice \cite{MatsubaraM1956}.
In the following we take $J$ as the energy unit and define a reduced temperature $t = T/J$ and a reduced field $h = H/J$ for convenience.
We investigate this model using numerically exact QMC simulations, based on the stochastic series expansion (SSE) algorithm~\cite{SyljuasenS2002}. By using a $\beta$-doubling procedure~\cite{Sandvik2002}, we simulate system sizes up to $L^3=26^3$ and with temperatures down to $t = 10^{-3}$. Our results are averaged over $2400$-$5300$ disorder realizations (depending on the size). Compared to previous works on $S=1$ disordered spin models by some of us~\cite{Yuetal2012, Yuetal2012-b}, the system studied here are at least twice as large, and the temperature (in units of the interaction) are lower than all previous studies to our knowledge, while the disorder statistics is of the same order.

 We characterize the equilibrium behavior of the system via the transverse static structure factor at wavevector ${\bm q}$,
${\mathcal{S}^{xy} }(\bm{q}) = {L^{-3}} \sum_{i,j} {e^{i\bm{q}\cdot( {{\bm{r}_i} - {{\bm{r}_j}} )}}\left\langle {S_i^xS_j^x + S_i^yS_j^y} \right\rangle}$
which is proportional to the momentum distribution of the hardcore bosons.  Such a quantity allows us to define the squared order parameter characterizing the long-range antiferromagnetic phase in the XY plane as $m_s^2=\mathcal{S}^{xy}(\bm{Q})/L^3$, where $\bm{Q}=(\pi,\pi,\pi)$ is the ordering wavevector; this quantity is proportional to the condensate density for bosons. A finite value of $m_s^2$ in the thermodynamic limit marks the phase with long-range antiferromagnetic order in the XY plane (AF-XY), which maps onto the SF phase for bosons.
The correlation length can be determined by using the second-moment estimator 
$ \xi = \frac{L}{{2\pi }}\sqrt {\frac{{\mathcal{S}^{xy}(\bm{Q})}}{{\mathcal{S}^{xy}(\bm{Q} + (2\pi/L,0,0)}} - 1}~.$
Throughout our work, the critical temperatures $t_c(x,h)$  have been determined via the crossing points of $\xi/L$ curves for different values of $L$, typically at fixed $t$ and variable $x$ or $h$.

The ground-state phase diagram, obtained via the $\xi/L$ scaling at the lowest temperature $t=10^{-3}$, is shown in Fig.~\ref{fig:1} in the dilution-vs-field plane. Two fundamental features can be easily established: 1) at the percolation threshold  $x_p = 0.6883923(2)$ \cite{Wangetal2013}, the site-diluted cubic lattice breaks up into finite clusters, so that long-range magnetic order is no longer sustainable for $x>x_p$; and 2) the clean system ($x=0$) undergoes a polarization transition at the saturation field $h_c(x=0) = 6$. Adding any finite amount of disorder to the system in the form of dilution alters the latter transition, introducing an intermediate BG phase \cite{Polletetal2009},  which necessarily erodes the SF phase, as the polarizing field is independent of the dilution.
On the other hand, in zero field a magnetic percolation transition can occur, which is purely driven by the geometry of the underlying lattice. Such a transition persists up to a multicritical field $h^*\approx 1$, at which the SF-BG transition line and the magnetic percolation line meet.

\begin{figure}
	\includegraphics[width=8.5cm]{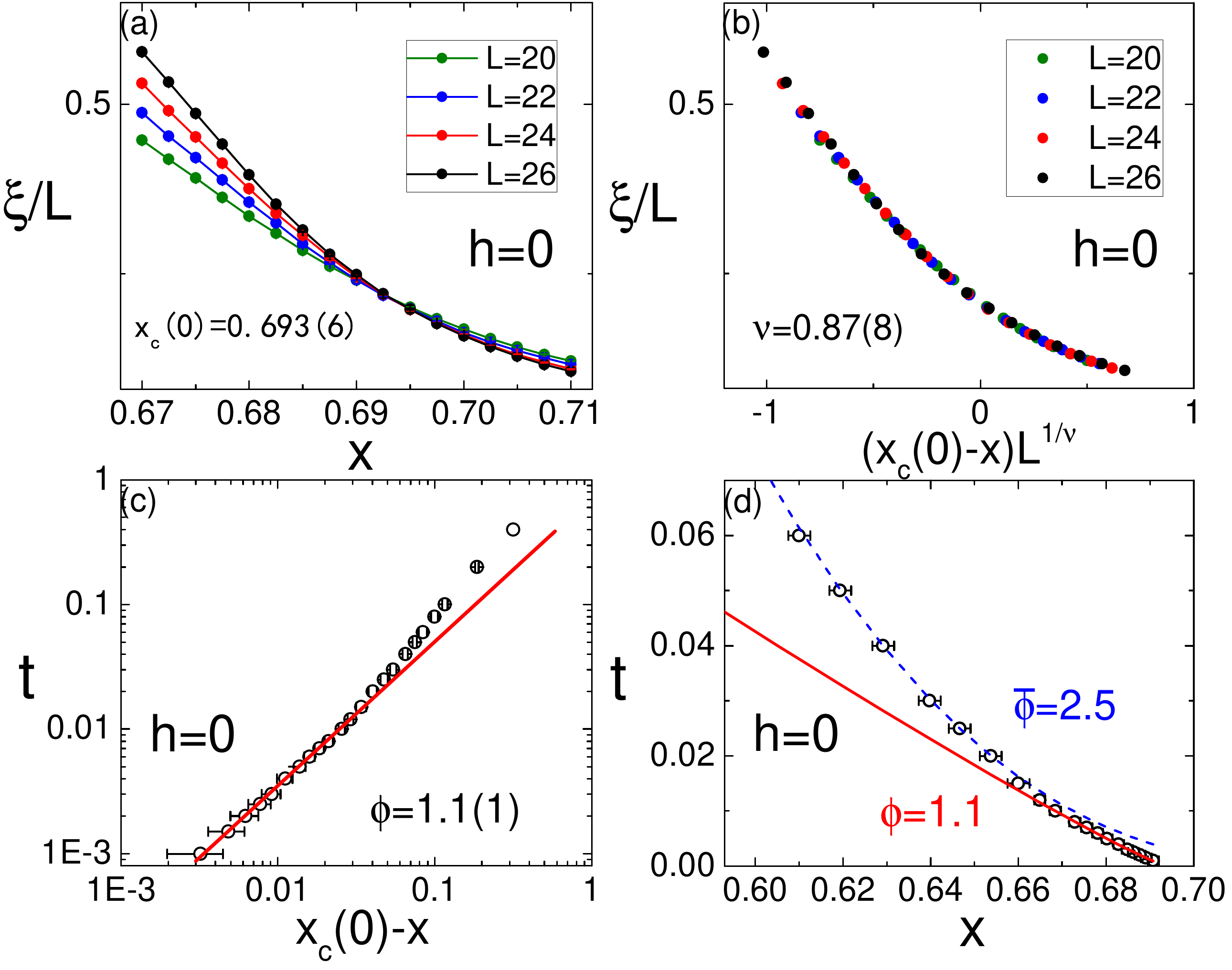} 
	\caption{Scaling at the zero-field magnetic transition. (a) and (b): Finite-size scaling of the correlation length $\xi$ at $t=0.001$, leading to the estimates $x_c=0.693(6)$ and $\nu=0.87(8)$. (c) and (d): Scaling of the finite-temperature transition line in the log-log (in (c)) and normal (in (d)) scales, respectively. Solid red lines are power-law fit $x_c = x_c(0) - At^{1/\phi}$ with $x_c(0)=0.694(6)$ and $\phi =1.1(1)$. The dashed blue line in panel (d) is a power-law fit with $\bar{\phi}=2.5$ to the intermediate temperature data, giving   $\bar{x}_c(0)\approx 0.732$.}
	\label{fig:2}
\end{figure}

\textit{Zero-field magnetic transition and percolation.} We first focus on the zero-field transition induced by dilution $x$. Upon diluting the cubic lattice one may wonder whether quantum fluctuations are able to destroy long-range order for $x<x_p$, but this turns out not to be the case -- as clearly shown in Fig.~\ref{fig:2}(a). There the scaling analysis of the correlation length at the lowest temperature ($t= 10^{-3}$) indicates a critical point of 0.693(6), consistent with $x_p$ (see below for the $t=0$ extrapolation). Moreover the estimated correlation length exponent $\nu = 0.87(8)$ -- obtained from the collapse of different data sets according to the scaling Ansatz $\xi/L = F(|x-x_c| L^{1/\nu})$ (Fig.~\ref{fig:2}(b)) -- is also found to be consistent with the percolation value $\nu_p = 0.876(1)$. As shown in the Supplemental Material (SM) \cite{SM}, adding corrections to scaling allows us to collapse the $\xi/L$ curves by using $x_c = x_p$, as well as to collapse curves for  $m_s^2$ with the 3d percolation exponent $\beta_p = 0.418(1)$ \cite{Wangetal2013}. We conclude that corrections to scaling, while necessary for the correct estimate of quantum critical exponents, only lead to a small ($\lesssim 1 \%$) shift in the estimate of the critical point. Therefore, in order to minimize the number of fitting factors, we shall not apply them in the rest of the work.
Our data are therefore fully consistent with a magnetic percolation transition occurring at $x_c = x_p$.
These results generalize to three dimensions the ones (both theoretical \cite{Sandvik2002} as well as experimental \cite{Vajketal2002}) already obtained for the magnetic percolation transition of the 2d version of this model.

The critical dilution $x_c(t)$ for the 3d Heisenberg model at various finite temperatures can be systematically determined via the $\xi/L$ scaling, and it is shown in Fig.~\ref{fig:2}(c-d). It clearly exhibits a power-law behavior  $x_c(t) = x_c(0) - A t^{1/\phi}$, stabilizing only in the lowest temperature range we explored ($t \sim 10^{-3} \div 10^{-2}$). Fitting the data in this range delivers $x_c(0) = 0.694(6)$, deviating by  $< 1\%$ from $x_p$; and $\phi = 1.1(1)$. Most interestingly, a similar value for $\phi$ has been recently obtained in an extensive numerical study on the classical ($S=\infty$) Heisenberg model on the same diluted lattice \cite{Khairnaretal2021}; and it has been shown to be consistent with the crossover exponent for percolation, $\phi = \nu_p \tilde\zeta_R = 1.12(2)$, where $\tilde\zeta_R = 1.28(2)$ dictates the scaling of the resistance of a resistor network defined on the critical percolating cluster \cite{Coniglio1981}. The coincidence of the behavior of the quantum system and that of the classical system is not at all surprising, given that quantum effects are found not to change the percolative nature of the magnetic transition at zero temperature; and they are even less expected to affect the finite-temperature behavior. On the other hand we remark that an exponent $\phi\geq 2$ could be easily fitted to our results when discarding the lowest temperature results; but this approach would lead to an unphysical estimate of the zero-temperature transition at $x_c(0) > x_p$.

\begin{figure}
	\includegraphics[width=8.5cm]{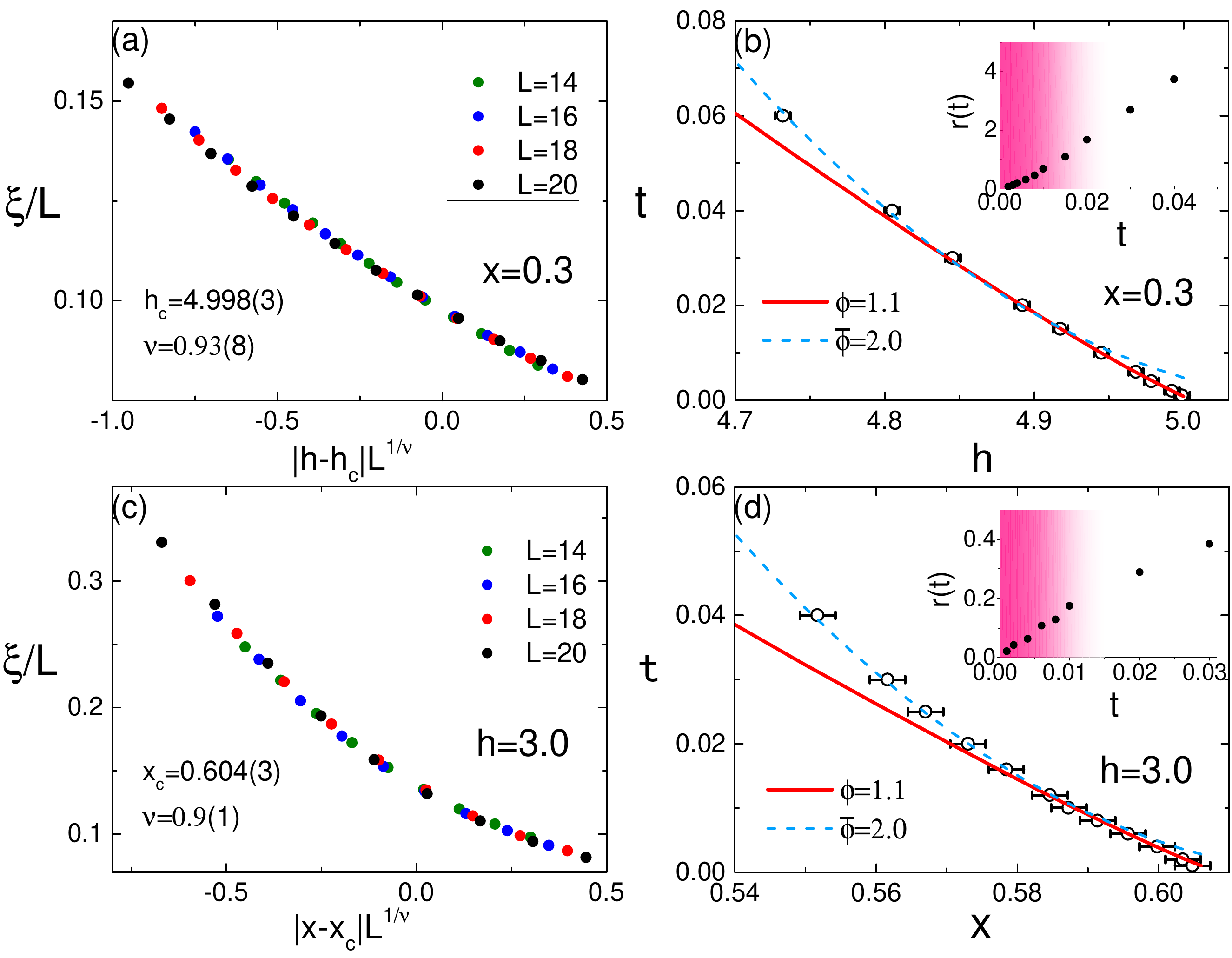} 
	\caption{Scaling at the SF-BG transition. (a,c) Scaling plots of the correlation length $\xi/L$ at $t=0.001$ and the finite-temperature transition line by varying $h$ at $x = 0.3$ (a) and by varying $x$ at $h=3$ (c). (b,d) Temperature-dependent critical points $h_c(t)$ at $x=0.3$ (b) and $x_c(t)$ data at $h=3$ (d). The red lines show fits to the power-law behavior $g_c(t) = g_c(0) -A t^{1/\phi}$ ($g=h,x$). The blue lines are fits of the intermediate temperature data with fixed $\bar \phi = 2$. Insets of (b,d): parameter $r(t)$ (see text) along the critical line; the pink-shaded region shows the temperature regime in which we observe $\phi\approx 1.1$.}
	\label{fig:3}
\end{figure}

\textit{SF-BG transition and $\phi$ exponent}.  We now move to the central part of the work, namely the study of the crossover exponent $\phi$ at the SF-BG transition.
To demonstrate the universality, we study the transition along two cuts of the SF-BG transition line: i) varying $h$ at $x=0.3$, and ii) varying $x$ at $h=3$.
The critical field and dilution concentration for the lowest accessible temperature $t=10^{-3}$ are determined from the scaling analysis of $\xi/L$ to be $h_c(t=10^{-3},x=0.3)=4.998(3)$ and $x_c(t=10^{-3},h=3)=0.604(3)$, respectively, with an estimated $\nu=0.9(1)$  -- see Fig.~\ref{fig:3}(a),(c).
The critical fields ($h_c(t)$) and concentrations ($x_c(t)$) at finite temperature, obtained with a similar scaling analysis, are shown in Figs.~\ref{fig:3}(b), (d), and fitted to
the power-law dependence $g_c(t) = g_c(0) - A t^{1/\phi}$ ($g=h,x$). Our data clearly indicate that $\phi = 1.1(1)$ in both cases -- a detailed fitting analysis on variable temperature windows is presented in the SM \cite{SM}. The curve best fitting the points at the lowest temperatures $t = 10^{-3}\div 10^{-2}$ has nearly zero curvature (since $\phi$ is very close to unity). Yet the prediction of conventional scaling theory ($\phi\geq 2$) would instead require a finite curvature, which can be easily found by looking at higher temperatures. A  $\phi=2$ exponent can fit our data only if we \textit{exclude} the points in the above-cited low-$t$ range, leading to extrapolated $t=0$ critical points
which are inconsistent with those ($h_c(t=0,x=0.3) =  5.007(3)$ and $x_c(t=0,h=3) = 0.606(5)$) obtained by using $\phi$ as a fitting parameter.

\begin{figure}
	\includegraphics[width=8.5cm]{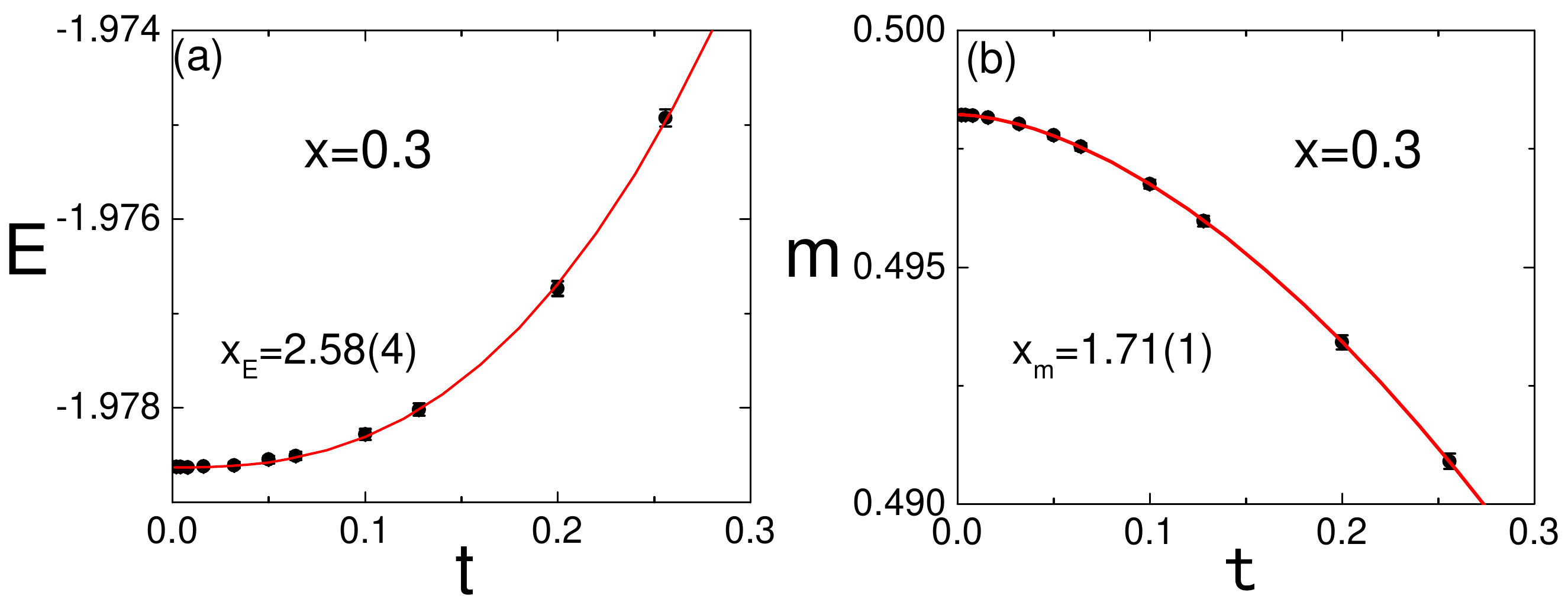} 
	\caption{Temperature evolution of the thermal energy per spin $E(t)$ (a) and the uniform magnetization $m(t)$ (b) along the QC trajectory at $x=0.3$, $h_c(0)=5.007$. The solid lines are power-law fits giving $x_E=2.58(4)$ and $x_m=1.71(1)$, respectively.  
}
	\label{fig:4}
\end{figure}

In addition to the analysis of the finite-temperature transition lines in the vicinity of the QCP, a completely alternative approach can be used to extract the $\phi$ exponent, based on the thermodynamic behavior along the so-called quantum critical (QC) trajectory, namely varying $t$ at fixed $g=g_c(0)$. Along this trajectory the specific heat scales as $C(t) \sim t^{x_C}$, and the uniform magnetization behaves as $m(t)-m(0) \sim t^{x_m}$, where $t_C$ and $t_m$ are the associated scaling exponents. It has been shown that the relation $\phi^{-1} = x_C - x_m + 1$
holds for all scaling Ans\"atze proposed so far for the transition~\cite{Yuetal2012-b}, including that of Ref.~\cite{Fisheretal1989}. The $x_C$ exponent can be most conveniently obtained via QMC from that governing the scaling of the thermal energy, $E(t)-E(0) \sim t^{x_E}$ with $x_E=x_C+1$. Data of $E(t)$ and $m(t)$ at $x=0.3$ and $h_c(0)=5.007$ are shown in Fig.~\ref{fig:4}(a) and (b), respectively. From the power-law fit, we obtain $x_E\approx 2.58(4)$ and $x_m=1.71(1)$, which leads to $\phi=1.20(7)$, fully consistent with the analysis based on the transition lines.

We notice that the density factor $r(t) = |n(t)/n_c-1|$ \cite{footnote} can take widely different values over the temperature ranges that consistently lead us to the estimate $\phi \approx 1.1$ --  $r \lesssim 1$ in Fig.~\ref{fig:3}(b), $r \lesssim 0.2$ Fig.~\ref{fig:3}(d), and {$r \lesssim 4$} in the temperature range $t \leq 0.25$ exhibiting the QC trajectory behavior in Fig.~\ref{fig:4}(a,b). These results question the relevance of the density criterion $r\ll 1$ put forward by Ref.~\cite{Yaoetal2014}.

\textit{Discussion and conclusions.} Using two independent approaches for the extraction of the crossover exponent at the SF-BG transition (via the vanishing of $T_c$; and via the thermodynamics along the QC trajectory) we unambiguously find a value of $\phi$ which violates the prediction from conventional scaling theory, $\phi \geq 2$.
On the other hand, our findings are consistent with experimental results on Br-DTN, as well as with extensive numerical studies of its model Hamiltonian  ~\cite{Yuetal2012, Yuetal2014}. One could in principle argue that our results may not sit at sufficiently low temperatures so as to observe the onset of the true quantum critical behavior. Yet the temperatures at which we pushed our calculations ($t=10^{-3}$) show clear signs of saturation of the thermodynamic quantities to their $t=0$ value (see data in the SM \cite{SM}), and correspondingly the critical values $x_c(t=10^{-3})$ and $h_c(t=10^{-3})$ for dilution and field are typically compatible with their extrapolated $t=0$ value within the error bar. Therefore we believe that our results reach a rather satisfactory level of control on thermal effects, at least up to the system sizes we could access. The results of Ref.~\cite{Yaoetal2014}, concluding that $\phi \approx 2.7$, have full control on the position of the $t=0$ QCP only for the disordered link-current model: remarkably this model has an additional (particle-hole) symmetry with respect to generic bosonic models with disorder. In the disordered hardcore-boson models studied by Ref.~\cite{Yaoetal2014} (not possessing this symmetry) the position of the QCP is not evaluated directly, and one cannot conclude whether or not the asymptotic scaling regime was reached. As we have shown above, curves with $\phi \geq 2$ can be fitted to our data as well in an intermediate temperature range ($t \gtrsim 10^{-2}$) which is the one reached by Ref.~\cite{Yaoetal2014} for hardcore bosons; but they are no longer working when going to even lower temperatures.

Our data show that the crossover $\phi$ exponent at the SF-BG transition is compatible with the one of 3d percolation, suggesting a link between the two phenomena, as proposed recently by Ref.~\cite{SyromyatnikovS2017}. Nonetheless, the link is rather complex, as the SF-BG quantum critical exponents are markedly different from the ones of percolation. As shown in the SM \cite{SM}, a scaling analysis of $m_s^2 L^{2\beta/\nu}$ at the SF-BG transition for $x=0.3$ leads to estimating the two critical exponents as $\nu=0.9(1)$ and $\beta = 1.2(1)$: while the first exponent is compatible with $\nu_p$ of 3d percolation, the second one is definitely not compatible with $\beta_p$, even after taking into account corrections to scaling. Hence a more complex scenario of quantum percolation \cite{Yuetal2010} should be in order, by which quantum fluctuations break up the lattice into uncorrelated domains as geometric percolation would; but the order-parameter correlations have a different spatial structure with respect to a system of geometrically disconnected clusters. We also point out that the physics we investigated in this work is expected to be unchanged upon changing the lattice geometry, as long as one considers a lattice supporting long-range $xy$ order in the absence of dilution and in an applied magnetic field. In this respect the SF-BG transition we investigated can be potentially realized in any magnetic insulator realizing 3d Heisenberg antiferromagnetism (for $S=1/2$ as well as higher spins), and subject to chemical doping of its magnetic ions -- several examples of such systems (especially layered antiferromagnets exhibiting a 3d N\'eel transition) exist in the literature, e.g. La$_2$Cu$_{1-x}$(Zn,Mg)$_x$O$_4$ \cite{Vajketal2002},
Rb$_2$Mn$_{1-x}$Mg$_x$F$_4$ \cite{Cowleyetal1993},    Mn$_{1-x}$Zn$_x$PS$_3$ \cite{Muldersetal2002}. Our predictions are therefore testable via the high-field and low-temperature properties of such materials, as well as similar ones.

\acknowledgements
\textit{Acknowledgements.}  This work has in part been supported by
the National Science Foundation of China Grant No. 12174441, Ministry of Science and Technology of China,
National Program on Key Research Project Grant No.2016YFA0300504 and Research Funds of Remnin University
of China Grant No. 18XNLG24 (RY). TR is supported by ANR (`EELS' project) and QuantERA (`MAQS' project).
R.Y. acknowledges the hospitality of ENS de Lyon.

\bibliography{biblio}

\begin{thebibliography}{42}%
\makeatletter
\providecommand \@ifxundefined [1]{%
 \@ifx{#1\undefined}
}%
\providecommand \@ifnum [1]{%
 \ifnum #1\expandafter \@firstoftwo
 \else \expandafter \@secondoftwo
 \fi
}%
\providecommand \@ifx [1]{%
 \ifx #1\expandafter \@firstoftwo
 \else \expandafter \@secondoftwo
 \fi
}%
\providecommand \natexlab [1]{#1}%
\providecommand \enquote  [1]{``#1''}%
\providecommand \bibnamefont  [1]{#1}%
\providecommand \bibfnamefont [1]{#1}%
\providecommand \citenamefont [1]{#1}%
\providecommand \href@noop [0]{\@secondoftwo}%
\providecommand \href [0]{\begingroup \@sanitize@url \@href}%
\providecommand \@href[1]{\@@startlink{#1}\@@href}%
\providecommand \@@href[1]{\endgroup#1\@@endlink}%
\providecommand \@sanitize@url [0]{\catcode `\\12\catcode `\$12\catcode
  `\&12\catcode `\#12\catcode `\^12\catcode `\_12\catcode `\%12\relax}%
\providecommand \@@startlink[1]{}%
\providecommand \@@endlink[0]{}%
\providecommand \url  [0]{\begingroup\@sanitize@url \@url }%
\providecommand \@url [1]{\endgroup\@href {#1}{\urlprefix }}%
\providecommand \urlprefix  [0]{URL }%
\providecommand \Eprint [0]{\href }%
\providecommand \doibase [0]{http://dx.doi.org/}%
\providecommand \selectlanguage [0]{\@gobble}%
\providecommand \bibinfo  [0]{\@secondoftwo}%
\providecommand \bibfield  [0]{\@secondoftwo}%
\providecommand \translation [1]{[#1]}%
\providecommand \BibitemOpen [0]{}%
\providecommand \bibitemStop [0]{}%
\providecommand \bibitemNoStop [0]{.\EOS\space}%
\providecommand \EOS [0]{\spacefactor3000\relax}%
\providecommand \BibitemShut  [1]{\csname bibitem#1\endcsname}%
\let\auto@bib@innerbib\@empty
\bibitem [{\citenamefont {Giamarchi}\ and\ \citenamefont
  {Schulz}(1987)}]{GiamarchiS1987}%
  \BibitemOpen
  \bibfield  {author} {\bibinfo {author} {\bibfnamefont {T.}~\bibnamefont
  {Giamarchi}}\ and\ \bibinfo {author} {\bibfnamefont {H.~J.}\ \bibnamefont
  {Schulz}},\ }\href {\doibase 10.1209/0295-5075/3/12/007} {\bibfield
  {journal} {\bibinfo  {journal} {Europhysics Letters ({EPL})}\ }\textbf
  {\bibinfo {volume} {3}},\ \bibinfo {pages} {1287} (\bibinfo {year}
  {1987})}\BibitemShut {NoStop}%
\bibitem [{\citenamefont {Fisher}\ \emph {et~al.}(1989)\citenamefont {Fisher},
  \citenamefont {Weichman}, \citenamefont {Grinstein},\ and\ \citenamefont
  {Fisher}}]{Fisheretal1989}%
  \BibitemOpen
  \bibfield  {author} {\bibinfo {author} {\bibfnamefont {M.~P.~A.}\
  \bibnamefont {Fisher}}, \bibinfo {author} {\bibfnamefont {P.~B.}\
  \bibnamefont {Weichman}}, \bibinfo {author} {\bibfnamefont {G.}~\bibnamefont
  {Grinstein}}, \ and\ \bibinfo {author} {\bibfnamefont {D.~S.}\ \bibnamefont
  {Fisher}},\ }\href {\doibase 10.1103/PhysRevB.40.546} {\bibfield  {journal}
  {\bibinfo  {journal} {Phys. Rev. B}\ }\textbf {\bibinfo {volume} {40}},\
  \bibinfo {pages} {546} (\bibinfo {year} {1989})}\BibitemShut {NoStop}%
\bibitem [{\citenamefont {Svistunov}\ \emph {et~al.}(2015)\citenamefont
  {Svistunov}, \citenamefont {Babaev},\ and\ \citenamefont
  {Prokof'ev}}]{Prokofevbook}%
  \BibitemOpen
  \bibfield  {author} {\bibinfo {author} {\bibfnamefont {B.~V.}\ \bibnamefont
  {Svistunov}}, \bibinfo {author} {\bibfnamefont {E.~S.}\ \bibnamefont
  {Babaev}}, \ and\ \bibinfo {author} {\bibfnamefont {N.~V.}\ \bibnamefont
  {Prokof'ev}},\ }\href@noop {} {\emph {\bibinfo {title} {Superfluid States of
  Matter}}}\ (\bibinfo  {publisher} {CRC press},\ \bibinfo {year}
  {2015})\BibitemShut {NoStop}%
\bibitem [{\citenamefont {Abrahams}(2010)}]{Abrahamsbook}%
  \BibitemOpen
  \bibinfo {editor} {\bibfnamefont {E.}~\bibnamefont {Abrahams}},\ ed.,\
  \href@noop {} {\emph {\bibinfo {title} {50 Years of Anderson Localization}}}\
  (\bibinfo  {publisher} {World Scientific},\ \bibinfo {year}
  {2010})\BibitemShut {NoStop}%
\bibitem [{\citenamefont {Dobrosavljevic}\ \emph {et~al.}(2010)\citenamefont
  {Dobrosavljevic}, \citenamefont {Trivedi},\ and\ \citenamefont
  {Valles~Jr.}}]{Dobrosavljevicbook}%
  \BibitemOpen
  \bibfield  {author} {\bibinfo {author} {\bibfnamefont {V.}~\bibnamefont
  {Dobrosavljevic}}, \bibinfo {author} {\bibfnamefont {N.}~\bibnamefont
  {Trivedi}}, \ and\ \bibinfo {author} {\bibfnamefont {J.~M.}\ \bibnamefont
  {Valles~Jr.}},\ }\href@noop {} {\emph {\bibinfo {title} {50 Years of Anderson
  Localization}}}\ (\bibinfo  {publisher} {World Scientific},\ \bibinfo {year}
  {2010})\BibitemShut {NoStop}%
\bibitem [{\citenamefont {Vojta}(2019)}]{Vojta2019}%
  \BibitemOpen
  \bibfield  {author} {\bibinfo {author} {\bibfnamefont {T.}~\bibnamefont
  {Vojta}},\ }\href {\doibase 10.1146/annurev-conmatphys-031218-013433}
  {\bibfield  {journal} {\bibinfo  {journal} {Annual Review of Condensed Matter
  Physics}\ }\textbf {\bibinfo {volume} {10}},\ \bibinfo {pages} {233}
  (\bibinfo {year} {2019})},\ \Eprint
  {http://arxiv.org/abs/https://doi.org/10.1146/annurev-conmatphys-031218-013433}
  {https://doi.org/10.1146/annurev-conmatphys-031218-013433} \BibitemShut
  {NoStop}%
\bibitem [{\citenamefont {Alet}\ and\ \citenamefont
  {Laflorencie}(2018)}]{AletL2018}%
  \BibitemOpen
  \bibfield  {author} {\bibinfo {author} {\bibfnamefont {F.}~\bibnamefont
  {Alet}}\ and\ \bibinfo {author} {\bibfnamefont {N.}~\bibnamefont
  {Laflorencie}},\ }\href {\doibase https://doi.org/10.1016/j.crhy.2018.03.003}
  {\bibfield  {journal} {\bibinfo  {journal} {Comptes Rendus Physique}\
  }\textbf {\bibinfo {volume} {19}},\ \bibinfo {pages} {498} (\bibinfo {year}
  {2018})}\BibitemShut {NoStop}%
\bibitem [{\citenamefont {Abanin}\ \emph {et~al.}(2019)\citenamefont {Abanin},
  \citenamefont {Altman}, \citenamefont {Bloch},\ and\ \citenamefont
  {Serbyn}}]{Abaninetal2019}%
  \BibitemOpen
  \bibfield  {author} {\bibinfo {author} {\bibfnamefont {D.~A.}\ \bibnamefont
  {Abanin}}, \bibinfo {author} {\bibfnamefont {E.}~\bibnamefont {Altman}},
  \bibinfo {author} {\bibfnamefont {I.}~\bibnamefont {Bloch}}, \ and\ \bibinfo
  {author} {\bibfnamefont {M.}~\bibnamefont {Serbyn}},\ }\href {\doibase
  10.1103/RevModPhys.91.021001} {\bibfield  {journal} {\bibinfo  {journal}
  {Rev. Mod. Phys.}\ }\textbf {\bibinfo {volume} {91}},\ \bibinfo {pages}
  {021001} (\bibinfo {year} {2019})}\BibitemShut {NoStop}%
\bibitem [{\citenamefont {Crowell}\ \emph {et~al.}(1997)\citenamefont
  {Crowell}, \citenamefont {Van~Keuls},\ and\ \citenamefont
  {Reppy}}]{Crowelletal1997}%
  \BibitemOpen
  \bibfield  {author} {\bibinfo {author} {\bibfnamefont {P.~A.}\ \bibnamefont
  {Crowell}}, \bibinfo {author} {\bibfnamefont {F.~W.}\ \bibnamefont
  {Van~Keuls}}, \ and\ \bibinfo {author} {\bibfnamefont {J.~D.}\ \bibnamefont
  {Reppy}},\ }\href {\doibase 10.1103/PhysRevB.55.12620} {\bibfield  {journal}
  {\bibinfo  {journal} {Phys. Rev. B}\ }\textbf {\bibinfo {volume} {55}},\
  \bibinfo {pages} {12620} (\bibinfo {year} {1997})}\BibitemShut {NoStop}%
\bibitem [{\citenamefont {D'Errico}\ \emph {et~al.}(2014)\citenamefont
  {D'Errico}, \citenamefont {Lucioni}, \citenamefont {Tanzi}, \citenamefont
  {Gori}, \citenamefont {Roux}, \citenamefont {McCulloch}, \citenamefont
  {Giamarchi}, \citenamefont {Inguscio},\ and\ \citenamefont
  {Modugno}}]{DErricoetal2014}%
  \BibitemOpen
  \bibfield  {author} {\bibinfo {author} {\bibfnamefont {C.}~\bibnamefont
  {D'Errico}}, \bibinfo {author} {\bibfnamefont {E.}~\bibnamefont {Lucioni}},
  \bibinfo {author} {\bibfnamefont {L.}~\bibnamefont {Tanzi}}, \bibinfo
  {author} {\bibfnamefont {L.}~\bibnamefont {Gori}}, \bibinfo {author}
  {\bibfnamefont {G.}~\bibnamefont {Roux}}, \bibinfo {author} {\bibfnamefont
  {I.~P.}\ \bibnamefont {McCulloch}}, \bibinfo {author} {\bibfnamefont
  {T.}~\bibnamefont {Giamarchi}}, \bibinfo {author} {\bibfnamefont
  {M.}~\bibnamefont {Inguscio}}, \ and\ \bibinfo {author} {\bibfnamefont
  {G.}~\bibnamefont {Modugno}},\ }\href {\doibase
  10.1103/PhysRevLett.113.095301} {\bibfield  {journal} {\bibinfo  {journal}
  {Phys. Rev. Lett.}\ }\textbf {\bibinfo {volume} {113}},\ \bibinfo {pages}
  {095301} (\bibinfo {year} {2014})}\BibitemShut {NoStop}%
\bibitem [{\citenamefont {Sac{\'e}p{\'e}}\ \emph {et~al.}(2011)\citenamefont
  {Sac{\'e}p{\'e}}, \citenamefont {Dubouchet}, \citenamefont {Chapelier},
  \citenamefont {Sanquer}, \citenamefont {Ovadia}, \citenamefont {Shahar},
  \citenamefont {Feigel'man},\ and\ \citenamefont {Ioffe}}]{Sacepeetal2011}%
  \BibitemOpen
  \bibfield  {author} {\bibinfo {author} {\bibfnamefont {B.}~\bibnamefont
  {Sac{\'e}p{\'e}}}, \bibinfo {author} {\bibfnamefont {T.}~\bibnamefont
  {Dubouchet}}, \bibinfo {author} {\bibfnamefont {C.}~\bibnamefont
  {Chapelier}}, \bibinfo {author} {\bibfnamefont {M.}~\bibnamefont {Sanquer}},
  \bibinfo {author} {\bibfnamefont {M.}~\bibnamefont {Ovadia}}, \bibinfo
  {author} {\bibfnamefont {D.}~\bibnamefont {Shahar}}, \bibinfo {author}
  {\bibfnamefont {M.}~\bibnamefont {Feigel'man}}, \ and\ \bibinfo {author}
  {\bibfnamefont {L.}~\bibnamefont {Ioffe}},\ }\href {\doibase
  10.1038/nphys1892} {\bibfield  {journal} {\bibinfo  {journal} {Nature
  Physics}\ }\textbf {\bibinfo {volume} {7}},\ \bibinfo {pages} {239} (\bibinfo
  {year} {2011})}\BibitemShut {NoStop}%
\bibitem [{\citenamefont {Zheludev}\ and\ \citenamefont
  {Roscilde}(2013)}]{ZheludevR2013}%
  \BibitemOpen
  \bibfield  {author} {\bibinfo {author} {\bibfnamefont {A.}~\bibnamefont
  {Zheludev}}\ and\ \bibinfo {author} {\bibfnamefont {T.}~\bibnamefont
  {Roscilde}},\ }\href {\doibase https://doi.org/10.1016/j.crhy.2013.10.001}
  {\bibfield  {journal} {\bibinfo  {journal} {Comptes Rendus Physique}\
  }\textbf {\bibinfo {volume} {14}},\ \bibinfo {pages} {740} (\bibinfo {year}
  {2013})},\ \bibinfo {note} {disordered systems / Systèmes
  désordonnés}\BibitemShut {NoStop}%
\bibitem [{\citenamefont {Sachdev}(2001)}]{Sachdevbook}%
  \BibitemOpen
  \bibfield  {author} {\bibinfo {author} {\bibfnamefont {S.}~\bibnamefont
  {Sachdev}},\ }\href@noop {} {\emph {\bibinfo {title} {Quantum Phase
  Transitions}}}\ (\bibinfo  {publisher} {Cambridge University Press},\
  \bibinfo {year} {2001})\BibitemShut {NoStop}%
\bibitem [{\citenamefont {Giamarchi}\ \emph {et~al.}(2008)\citenamefont
  {Giamarchi}, \citenamefont {R{\"u}egg},\ and\ \citenamefont
  {Tchernyshyov}}]{Giamarchietal2008}%
  \BibitemOpen
  \bibfield  {author} {\bibinfo {author} {\bibfnamefont {T.}~\bibnamefont
  {Giamarchi}}, \bibinfo {author} {\bibfnamefont {C.}~\bibnamefont
  {R{\"u}egg}}, \ and\ \bibinfo {author} {\bibfnamefont {O.}~\bibnamefont
  {Tchernyshyov}},\ }\href {\doibase 10.1038/nphys893} {\bibfield  {journal}
  {\bibinfo  {journal} {Nature Physics}\ }\textbf {\bibinfo {volume} {4}},\
  \bibinfo {pages} {198} (\bibinfo {year} {2008})}\BibitemShut {NoStop}%
\bibitem [{\citenamefont {Zapf}\ \emph {et~al.}(2014)\citenamefont {Zapf},
  \citenamefont {Jaime},\ and\ \citenamefont {Batista}}]{Zapfetal2014}%
  \BibitemOpen
  \bibfield  {author} {\bibinfo {author} {\bibfnamefont {V.}~\bibnamefont
  {Zapf}}, \bibinfo {author} {\bibfnamefont {M.}~\bibnamefont {Jaime}}, \ and\
  \bibinfo {author} {\bibfnamefont {C.~D.}\ \bibnamefont {Batista}},\ }\href
  {\doibase 10.1103/RevModPhys.86.563} {\bibfield  {journal} {\bibinfo
  {journal} {Rev. Mod. Phys.}\ }\textbf {\bibinfo {volume} {86}},\ \bibinfo
  {pages} {563} (\bibinfo {year} {2014})}\BibitemShut {NoStop}%
\bibitem [{\citenamefont {Manaka}\ \emph {et~al.}(2008)\citenamefont {Manaka},
  \citenamefont {Kolomiets},\ and\ \citenamefont {Goto}}]{Manakaetal2008}%
  \BibitemOpen
  \bibfield  {author} {\bibinfo {author} {\bibfnamefont {H.}~\bibnamefont
  {Manaka}}, \bibinfo {author} {\bibfnamefont {A.~V.}\ \bibnamefont
  {Kolomiets}}, \ and\ \bibinfo {author} {\bibfnamefont {T.}~\bibnamefont
  {Goto}},\ }\href {\doibase 10.1103/PhysRevLett.101.077204} {\bibfield
  {journal} {\bibinfo  {journal} {Phys. Rev. Lett.}\ }\textbf {\bibinfo
  {volume} {101}},\ \bibinfo {pages} {077204} (\bibinfo {year}
  {2008})}\BibitemShut {NoStop}%
\bibitem [{\citenamefont {Hong}\ \emph {et~al.}(2010)\citenamefont {Hong},
  \citenamefont {Zheludev}, \citenamefont {Manaka},\ and\ \citenamefont
  {Regnault}}]{Hongetal2010}%
  \BibitemOpen
  \bibfield  {author} {\bibinfo {author} {\bibfnamefont {T.}~\bibnamefont
  {Hong}}, \bibinfo {author} {\bibfnamefont {A.}~\bibnamefont {Zheludev}},
  \bibinfo {author} {\bibfnamefont {H.}~\bibnamefont {Manaka}}, \ and\ \bibinfo
  {author} {\bibfnamefont {L.-P.}\ \bibnamefont {Regnault}},\ }\href {\doibase
  10.1103/PhysRevB.81.060410} {\bibfield  {journal} {\bibinfo  {journal} {Phys.
  Rev. B}\ }\textbf {\bibinfo {volume} {81}},\ \bibinfo {pages} {060410}
  (\bibinfo {year} {2010})}\BibitemShut {NoStop}%
\bibitem [{\citenamefont {Yamada}\ \emph {et~al.}(2011)\citenamefont {Yamada},
  \citenamefont {Tanaka}, \citenamefont {Ono},\ and\ \citenamefont
  {Nojiri}}]{Yamadaetal2011}%
  \BibitemOpen
  \bibfield  {author} {\bibinfo {author} {\bibfnamefont {F.}~\bibnamefont
  {Yamada}}, \bibinfo {author} {\bibfnamefont {H.}~\bibnamefont {Tanaka}},
  \bibinfo {author} {\bibfnamefont {T.}~\bibnamefont {Ono}}, \ and\ \bibinfo
  {author} {\bibfnamefont {H.}~\bibnamefont {Nojiri}},\ }\href {\doibase
  10.1103/PhysRevB.83.020409} {\bibfield  {journal} {\bibinfo  {journal} {Phys.
  Rev. B}\ }\textbf {\bibinfo {volume} {83}},\ \bibinfo {pages} {020409}
  (\bibinfo {year} {2011})}\BibitemShut {NoStop}%
\bibitem [{\citenamefont {H\"uvonen}\ \emph {et~al.}(2012)\citenamefont
  {H\"uvonen}, \citenamefont {Zhao}, \citenamefont {M\aa{}nsson}, \citenamefont
  {Yankova}, \citenamefont {Ressouche}, \citenamefont {Niedermayer},
  \citenamefont {Laver}, \citenamefont {Gvasaliya},\ and\ \citenamefont
  {Zheludev}}]{Huvonenetal2012}%
  \BibitemOpen
  \bibfield  {author} {\bibinfo {author} {\bibfnamefont {D.}~\bibnamefont
  {H\"uvonen}}, \bibinfo {author} {\bibfnamefont {S.}~\bibnamefont {Zhao}},
  \bibinfo {author} {\bibfnamefont {M.}~\bibnamefont {M\aa{}nsson}}, \bibinfo
  {author} {\bibfnamefont {T.}~\bibnamefont {Yankova}}, \bibinfo {author}
  {\bibfnamefont {E.}~\bibnamefont {Ressouche}}, \bibinfo {author}
  {\bibfnamefont {C.}~\bibnamefont {Niedermayer}}, \bibinfo {author}
  {\bibfnamefont {M.}~\bibnamefont {Laver}}, \bibinfo {author} {\bibfnamefont
  {S.~N.}\ \bibnamefont {Gvasaliya}}, \ and\ \bibinfo {author} {\bibfnamefont
  {A.}~\bibnamefont {Zheludev}},\ }\href {\doibase 10.1103/PhysRevB.85.100410}
  {\bibfield  {journal} {\bibinfo  {journal} {Phys. Rev. B}\ }\textbf {\bibinfo
  {volume} {85}},\ \bibinfo {pages} {100410} (\bibinfo {year}
  {2012})}\BibitemShut {NoStop}%
\bibitem [{\citenamefont {Yu}\ \emph {et~al.}(2012{\natexlab{a}})\citenamefont
  {Yu}, \citenamefont {Yin}, \citenamefont {Sullivan}, \citenamefont {Xia},
  \citenamefont {Huan}, \citenamefont {Paduan-Filho}, \citenamefont
  {Oliveira~Jr}, \citenamefont {Haas}, \citenamefont {Steppke}, \citenamefont
  {Miclea}, \citenamefont {Weickert}, \citenamefont {Movshovich}, \citenamefont
  {Mun}, \citenamefont {Scott}, \citenamefont {Zapf},\ and\ \citenamefont
  {Roscilde}}]{Yuetal2012}%
  \BibitemOpen
  \bibfield  {author} {\bibinfo {author} {\bibfnamefont {R.}~\bibnamefont
  {Yu}}, \bibinfo {author} {\bibfnamefont {L.}~\bibnamefont {Yin}}, \bibinfo
  {author} {\bibfnamefont {N.~S.}\ \bibnamefont {Sullivan}}, \bibinfo {author}
  {\bibfnamefont {J.~S.}\ \bibnamefont {Xia}}, \bibinfo {author} {\bibfnamefont
  {C.}~\bibnamefont {Huan}}, \bibinfo {author} {\bibfnamefont {A.}~\bibnamefont
  {Paduan-Filho}}, \bibinfo {author} {\bibfnamefont {N.~F.}\ \bibnamefont
  {Oliveira~Jr}}, \bibinfo {author} {\bibfnamefont {S.}~\bibnamefont {Haas}},
  \bibinfo {author} {\bibfnamefont {A.}~\bibnamefont {Steppke}}, \bibinfo
  {author} {\bibfnamefont {C.~F.}\ \bibnamefont {Miclea}}, \bibinfo {author}
  {\bibfnamefont {F.}~\bibnamefont {Weickert}}, \bibinfo {author}
  {\bibfnamefont {R.}~\bibnamefont {Movshovich}}, \bibinfo {author}
  {\bibfnamefont {E.-D.}\ \bibnamefont {Mun}}, \bibinfo {author} {\bibfnamefont
  {B.~L.}\ \bibnamefont {Scott}}, \bibinfo {author} {\bibfnamefont {V.~S.}\
  \bibnamefont {Zapf}}, \ and\ \bibinfo {author} {\bibfnamefont
  {T.}~\bibnamefont {Roscilde}},\ }\href {\doibase 10.1038/nature11406}
  {\bibfield  {journal} {\bibinfo  {journal} {Nature}\ }\textbf {\bibinfo
  {volume} {489}},\ \bibinfo {pages} {379} (\bibinfo {year}
  {2012}{\natexlab{a}})}\BibitemShut {NoStop}%
\bibitem [{\citenamefont {Yu}\ \emph {et~al.}(2012{\natexlab{b}})\citenamefont
  {Yu}, \citenamefont {Miclea}, \citenamefont {Weickert}, \citenamefont
  {Movshovich}, \citenamefont {Paduan-Filho}, \citenamefont {Zapf},\ and\
  \citenamefont {Roscilde}}]{Yuetal2012-b}%
  \BibitemOpen
  \bibfield  {author} {\bibinfo {author} {\bibfnamefont {R.}~\bibnamefont
  {Yu}}, \bibinfo {author} {\bibfnamefont {C.~F.}\ \bibnamefont {Miclea}},
  \bibinfo {author} {\bibfnamefont {F.}~\bibnamefont {Weickert}}, \bibinfo
  {author} {\bibfnamefont {R.}~\bibnamefont {Movshovich}}, \bibinfo {author}
  {\bibfnamefont {A.}~\bibnamefont {Paduan-Filho}}, \bibinfo {author}
  {\bibfnamefont {V.~S.}\ \bibnamefont {Zapf}}, \ and\ \bibinfo {author}
  {\bibfnamefont {T.}~\bibnamefont {Roscilde}},\ }\href {\doibase
  10.1103/PhysRevB.86.134421} {\bibfield  {journal} {\bibinfo  {journal} {Phys.
  Rev. B}\ }\textbf {\bibinfo {volume} {86}},\ \bibinfo {pages} {134421}
  (\bibinfo {year} {2012}{\natexlab{b}})}\BibitemShut {NoStop}%
\bibitem [{\citenamefont {Yu}\ \emph {et~al.}(2010)\citenamefont {Yu},
  \citenamefont {Haas},\ and\ \citenamefont {Roscilde}}]{Yuetal2010}%
  \BibitemOpen
  \bibfield  {author} {\bibinfo {author} {\bibfnamefont {R.}~\bibnamefont
  {Yu}}, \bibinfo {author} {\bibfnamefont {S.}~\bibnamefont {Haas}}, \ and\
  \bibinfo {author} {\bibfnamefont {T.}~\bibnamefont {Roscilde}},\ }\href
  {\doibase 10.1209/0295-5075/89/10009} {\bibfield  {journal} {\bibinfo
  {journal} {{EPL} (Europhysics Letters)}\ }\textbf {\bibinfo {volume} {89}},\
  \bibinfo {pages} {10009} (\bibinfo {year} {2010})}\BibitemShut {NoStop}%
\bibitem [{\citenamefont {Hitchcock}\ and\ \citenamefont
  {S\o{}rensen}(2006)}]{HitchcockS2006}%
  \BibitemOpen
  \bibfield  {author} {\bibinfo {author} {\bibfnamefont {P.}~\bibnamefont
  {Hitchcock}}\ and\ \bibinfo {author} {\bibfnamefont {E.~S.}\ \bibnamefont
  {S\o{}rensen}},\ }\href {\doibase 10.1103/PhysRevB.73.174523} {\bibfield
  {journal} {\bibinfo  {journal} {Phys. Rev. B}\ }\textbf {\bibinfo {volume}
  {73}},\ \bibinfo {pages} {174523} (\bibinfo {year} {2006})}\BibitemShut
  {NoStop}%
\bibitem [{\citenamefont {Yao}\ \emph {et~al.}(2014)\citenamefont {Yao},
  \citenamefont {da~Costa}, \citenamefont {Kiselev},\ and\ \citenamefont
  {Prokof'ev}}]{Yaoetal2014}%
  \BibitemOpen
  \bibfield  {author} {\bibinfo {author} {\bibfnamefont {Z.}~\bibnamefont
  {Yao}}, \bibinfo {author} {\bibfnamefont {K.~P.~C.}\ \bibnamefont
  {da~Costa}}, \bibinfo {author} {\bibfnamefont {M.}~\bibnamefont {Kiselev}}, \
  and\ \bibinfo {author} {\bibfnamefont {N.}~\bibnamefont {Prokof'ev}},\ }\href
  {\doibase 10.1103/PhysRevLett.112.225301} {\bibfield  {journal} {\bibinfo
  {journal} {Phys. Rev. Lett.}\ }\textbf {\bibinfo {volume} {112}},\ \bibinfo
  {pages} {225301} (\bibinfo {year} {2014})}\BibitemShut {NoStop}%
\bibitem [{\citenamefont {Meier}\ and\ \citenamefont
  {Wallin}(2012)}]{MeierW2012}%
  \BibitemOpen
  \bibfield  {author} {\bibinfo {author} {\bibfnamefont {H.}~\bibnamefont
  {Meier}}\ and\ \bibinfo {author} {\bibfnamefont {M.}~\bibnamefont {Wallin}},\
  }\href {\doibase 10.1103/PhysRevLett.108.055701} {\bibfield  {journal}
  {\bibinfo  {journal} {Phys. Rev. Lett.}\ }\textbf {\bibinfo {volume} {108}},\
  \bibinfo {pages} {055701} (\bibinfo {year} {2012})}\BibitemShut {NoStop}%
\bibitem [{\citenamefont {Ng}\ and\ \citenamefont
  {S\o{}rensen}(2015)}]{NgS2015}%
  \BibitemOpen
  \bibfield  {author} {\bibinfo {author} {\bibfnamefont {R.}~\bibnamefont
  {Ng}}\ and\ \bibinfo {author} {\bibfnamefont {E.~S.}\ \bibnamefont
  {S\o{}rensen}},\ }\href {\doibase 10.1103/PhysRevLett.114.255701} {\bibfield
  {journal} {\bibinfo  {journal} {Phys. Rev. Lett.}\ }\textbf {\bibinfo
  {volume} {114}},\ \bibinfo {pages} {255701} (\bibinfo {year}
  {2015})}\BibitemShut {NoStop}%
\bibitem [{\citenamefont {Yu}\ \emph {et~al.}(2014)\citenamefont {Yu},
  \citenamefont {Zapf},\ and\ \citenamefont {Roscilde}}]{Yuetal2014}%
  \BibitemOpen
  \bibfield  {author} {\bibinfo {author} {\bibfnamefont {R.}~\bibnamefont
  {Yu}}, \bibinfo {author} {\bibfnamefont {V.}~\bibnamefont {Zapf}}, \ and\
  \bibinfo {author} {\bibfnamefont {T.}~\bibnamefont {Roscilde}},\ }\href
  {\doibase 10.48550/ARXIV.1403.6059} {\enquote {\bibinfo {title} {Comment to
  "critical exponents of the superfluid-bose glass transition in
  three-dimensions" by z. yao et al., arxiv:1402.5417v1},}\ } (\bibinfo {year}
  {2014})\BibitemShut {NoStop}%
\bibitem [{\citenamefont {Stauffer}\ and\ \citenamefont
  {Aharony}(1992)}]{percolationbook}%
  \BibitemOpen
  \bibfield  {author} {\bibinfo {author} {\bibfnamefont {D.}~\bibnamefont
  {Stauffer}}\ and\ \bibinfo {author} {\bibfnamefont {A.}~\bibnamefont
  {Aharony}},\ }\href@noop {} {\emph {\bibinfo {title} {Introduction to
  Percolation Theory}}}\ (\bibinfo  {publisher} {Taylor \& Francis},\ \bibinfo
  {year} {1992})\BibitemShut {NoStop}%
\bibitem [{\citenamefont {Coniglio}(1981)}]{Coniglio1981}%
  \BibitemOpen
  \bibfield  {author} {\bibinfo {author} {\bibfnamefont {A.}~\bibnamefont
  {Coniglio}},\ }\href {\doibase 10.1103/PhysRevLett.46.250} {\bibfield
  {journal} {\bibinfo  {journal} {Phys. Rev. Lett.}\ }\textbf {\bibinfo
  {volume} {46}},\ \bibinfo {pages} {250} (\bibinfo {year} {1981})}\BibitemShut
  {NoStop}%
\bibitem [{\citenamefont {{Khairnar, Gaurav}}\ \emph
  {et~al.}(2021)\citenamefont {{Khairnar, Gaurav}}, \citenamefont {{Lerch,
  Cameron}},\ and\ \citenamefont {{Vojta, Thomas}}}]{Khairnaretal2021}%
  \BibitemOpen
  \bibfield  {author} {\bibinfo {author} {\bibnamefont {{Khairnar, Gaurav}}},
  \bibinfo {author} {\bibnamefont {{Lerch, Cameron}}}, \ and\ \bibinfo {author}
  {\bibnamefont {{Vojta, Thomas}}},\ }\href {\doibase
  10.1140/epjb/s10051-021-00056-4} {\bibfield  {journal} {\bibinfo  {journal}
  {Eur. Phys. J. B}\ }\textbf {\bibinfo {volume} {94}},\ \bibinfo {pages} {43}
  (\bibinfo {year} {2021})}\BibitemShut {NoStop}%
\bibitem [{\citenamefont {Matsubara}\ and\ \citenamefont
  {Matsuda}(1956)}]{MatsubaraM1956}%
  \BibitemOpen
  \bibfield  {author} {\bibinfo {author} {\bibfnamefont {T.}~\bibnamefont
  {Matsubara}}\ and\ \bibinfo {author} {\bibfnamefont {H.}~\bibnamefont
  {Matsuda}},\ }\href {\doibase 10.1143/PTP.16.569} {\bibfield  {journal}
  {\bibinfo  {journal} {Progress of Theoretical Physics}\ }\textbf {\bibinfo
  {volume} {16}},\ \bibinfo {pages} {569} (\bibinfo {year} {1956})},\ \Eprint
  {http://arxiv.org/abs/https://academic.oup.com/ptp/article-pdf/16/6/569/5383838/16-6-569.pdf}
  {https://academic.oup.com/ptp/article-pdf/16/6/569/5383838/16-6-569.pdf}
  \BibitemShut {NoStop}%
\bibitem [{\citenamefont {Sylju\aa{}sen}\ and\ \citenamefont
  {Sandvik}(2002)}]{SyljuasenS2002}%
  \BibitemOpen
  \bibfield  {author} {\bibinfo {author} {\bibfnamefont {O.~F.}\ \bibnamefont
  {Sylju\aa{}sen}}\ and\ \bibinfo {author} {\bibfnamefont {A.~W.}\ \bibnamefont
  {Sandvik}},\ }\href {\doibase 10.1103/PhysRevE.66.046701} {\bibfield
  {journal} {\bibinfo  {journal} {Phys. Rev. E}\ }\textbf {\bibinfo {volume}
  {66}},\ \bibinfo {pages} {046701} (\bibinfo {year} {2002})}\BibitemShut
  {NoStop}%
\bibitem [{\citenamefont {Sandvik}(2002)}]{Sandvik2002}%
  \BibitemOpen
  \bibfield  {author} {\bibinfo {author} {\bibfnamefont {A.~W.}\ \bibnamefont
  {Sandvik}},\ }\href {\doibase 10.1103/PhysRevB.66.024418} {\bibfield
  {journal} {\bibinfo  {journal} {Phys. Rev. B}\ }\textbf {\bibinfo {volume}
  {66}},\ \bibinfo {pages} {024418} (\bibinfo {year} {2002})}\BibitemShut
  {NoStop}%
\bibitem [{\citenamefont {Wang}\ \emph {et~al.}(2013)\citenamefont {Wang},
  \citenamefont {Zhou}, \citenamefont {Zhang}, \citenamefont {Garoni},\ and\
  \citenamefont {Deng}}]{Wangetal2013}%
  \BibitemOpen
  \bibfield  {author} {\bibinfo {author} {\bibfnamefont {J.}~\bibnamefont
  {Wang}}, \bibinfo {author} {\bibfnamefont {Z.}~\bibnamefont {Zhou}}, \bibinfo
  {author} {\bibfnamefont {W.}~\bibnamefont {Zhang}}, \bibinfo {author}
  {\bibfnamefont {T.~M.}\ \bibnamefont {Garoni}}, \ and\ \bibinfo {author}
  {\bibfnamefont {Y.}~\bibnamefont {Deng}},\ }\href {\doibase
  10.1103/PhysRevE.87.052107} {\bibfield  {journal} {\bibinfo  {journal} {Phys.
  Rev. E}\ }\textbf {\bibinfo {volume} {87}},\ \bibinfo {pages} {052107}
  (\bibinfo {year} {2013})}\BibitemShut {NoStop}%
\bibitem [{\citenamefont {Pollet}\ \emph {et~al.}(2009)\citenamefont {Pollet},
  \citenamefont {Prokof'ev}, \citenamefont {Svistunov},\ and\ \citenamefont
  {Troyer}}]{Polletetal2009}%
  \BibitemOpen
  \bibfield  {author} {\bibinfo {author} {\bibfnamefont {L.}~\bibnamefont
  {Pollet}}, \bibinfo {author} {\bibfnamefont {N.~V.}\ \bibnamefont
  {Prokof'ev}}, \bibinfo {author} {\bibfnamefont {B.~V.}\ \bibnamefont
  {Svistunov}}, \ and\ \bibinfo {author} {\bibfnamefont {M.}~\bibnamefont
  {Troyer}},\ }\href {\doibase 10.1103/PhysRevLett.103.140402} {\bibfield
  {journal} {\bibinfo  {journal} {Phys. Rev. Lett.}\ }\textbf {\bibinfo
  {volume} {103}},\ \bibinfo {pages} {140402} (\bibinfo {year}
  {2009})}\BibitemShut {NoStop}%
\bibitem [{SM()}]{SM}%
  \BibitemOpen
  \href@noop {} {}\bibinfo {note} {See {S}upplemental {M}aterial ({SM}) for
  details about: 1) the convergence of our results to their zero-temperature
  limit; 2) the scaling analysis of the correlation length and magnetization in
  zero field, showing that a consistent scaling collapse (including corrections
  to scaling) can be obtained using critical exponents and critical point from
  3d percolation; 3) a windowing analysis of the extracted $\phi$ exponent from
  fits of the $t_c(h)$ and $t_c(x)$ data at the SF-BG transition; 4) a scaling
  analysis of $m_s^2$ at the SF-BG transition, showing that it is incompatible
  with percolation exponents.}\BibitemShut {Stop}%
\bibitem [{\citenamefont {Vajk}\ \emph {et~al.}(2002)\citenamefont {Vajk},
  \citenamefont {Mang}, \citenamefont {Greven}, \citenamefont {Gehring},\ and\
  \citenamefont {Lynn}}]{Vajketal2002}%
  \BibitemOpen
  \bibfield  {author} {\bibinfo {author} {\bibfnamefont {O.~P.}\ \bibnamefont
  {Vajk}}, \bibinfo {author} {\bibfnamefont {P.~K.}\ \bibnamefont {Mang}},
  \bibinfo {author} {\bibfnamefont {M.}~\bibnamefont {Greven}}, \bibinfo
  {author} {\bibfnamefont {P.~M.}\ \bibnamefont {Gehring}}, \ and\ \bibinfo
  {author} {\bibfnamefont {J.~W.}\ \bibnamefont {Lynn}},\ }\href {\doibase
  10.1126/science.1067110} {\bibfield  {journal} {\bibinfo  {journal}
  {Science}\ }\textbf {\bibinfo {volume} {295}},\ \bibinfo {pages} {1691}
  (\bibinfo {year} {2002})},\ \Eprint
  {http://arxiv.org/abs/https://www.science.org/doi/pdf/10.1126/science.1067110}
  {https://www.science.org/doi/pdf/10.1126/science.1067110} \BibitemShut
  {NoStop}%
\bibitem [{foo()}]{footnote}%
  \BibitemOpen
  \href@noop {} {}\bibinfo {note} {For this model, $n=1/2-m$ and hence
  $r=|n/n_c-1|=(m_c-m)/(1/2-m_c)$.}\BibitemShut {Stop}%
\bibitem [{\citenamefont {Syromyatnikov}\ and\ \citenamefont
  {Sizanov}(2017)}]{SyromyatnikovS2017}%
  \BibitemOpen
  \bibfield  {author} {\bibinfo {author} {\bibfnamefont {A.~V.}\ \bibnamefont
  {Syromyatnikov}}\ and\ \bibinfo {author} {\bibfnamefont {A.~V.}\ \bibnamefont
  {Sizanov}},\ }\href {\doibase 10.1103/PhysRevB.95.014206} {\bibfield
  {journal} {\bibinfo  {journal} {Phys. Rev. B}\ }\textbf {\bibinfo {volume}
  {95}},\ \bibinfo {pages} {014206} (\bibinfo {year} {2017})}\BibitemShut
  {NoStop}%
\bibitem [{\citenamefont {Cowley}\ \emph {et~al.}(1993)\citenamefont {Cowley},
  \citenamefont {Aharony}, \citenamefont {Birgeneau}, \citenamefont
  {Pelcovits}, \citenamefont {Shirane},\ and\ \citenamefont
  {Thurston}}]{Cowleyetal1993}%
  \BibitemOpen
  \bibfield  {author} {\bibinfo {author} {\bibfnamefont {R.~A.}\ \bibnamefont
  {Cowley}}, \bibinfo {author} {\bibfnamefont {A.}~\bibnamefont {Aharony}},
  \bibinfo {author} {\bibfnamefont {R.~J.}\ \bibnamefont {Birgeneau}}, \bibinfo
  {author} {\bibfnamefont {R.~A.}\ \bibnamefont {Pelcovits}}, \bibinfo {author}
  {\bibfnamefont {G.}~\bibnamefont {Shirane}}, \ and\ \bibinfo {author}
  {\bibfnamefont {T.~R.}\ \bibnamefont {Thurston}},\ }\href {\doibase
  10.1007/BF01308802} {\bibfield  {journal} {\bibinfo  {journal} {Zeitschrift
  f{\"u}r Physik B Condensed Matter}\ }\textbf {\bibinfo {volume} {93}},\
  \bibinfo {pages} {5} (\bibinfo {year} {1993})}\BibitemShut {NoStop}%
\bibitem [{\citenamefont {Mulders}\ \emph {et~al.}(2002)\citenamefont
  {Mulders}, \citenamefont {Klaasse}, \citenamefont {Goossens}, \citenamefont
  {Chadwick},\ and\ \citenamefont {Hicks}}]{Muldersetal2002}%
  \BibitemOpen
  \bibfield  {author} {\bibinfo {author} {\bibfnamefont {A.~M.}\ \bibnamefont
  {Mulders}}, \bibinfo {author} {\bibfnamefont {J.~C.~P.}\ \bibnamefont
  {Klaasse}}, \bibinfo {author} {\bibfnamefont {D.~J.}\ \bibnamefont
  {Goossens}}, \bibinfo {author} {\bibfnamefont {J.}~\bibnamefont {Chadwick}},
  \ and\ \bibinfo {author} {\bibfnamefont {T.~J.}\ \bibnamefont {Hicks}},\
  }\href {\doibase 10.1088/0953-8984/14/37/306} {\bibfield  {journal} {\bibinfo
   {journal} {Journal of Physics: Condensed Matter}\ }\textbf {\bibinfo
  {volume} {14}},\ \bibinfo {pages} {8697} (\bibinfo {year}
  {2002})}\BibitemShut {NoStop}%
\bibitem [{\citenamefont {Ballesteros}\ \emph {et~al.}(1999)\citenamefont
  {Ballesteros}, \citenamefont {Fern{\'{a}}ndez}, \citenamefont
  {Mart{\'{\i}}n-Mayor}, \citenamefont {Sudupe}, \citenamefont {Parisi},\ and\
  \citenamefont {Ruiz-Lorenzo}}]{Ballesterosetal1999}%
  \BibitemOpen
  \bibfield  {author} {\bibinfo {author} {\bibfnamefont {H.~G.}\ \bibnamefont
  {Ballesteros}}, \bibinfo {author} {\bibfnamefont {L.~A.}\ \bibnamefont
  {Fern{\'{a}}ndez}}, \bibinfo {author} {\bibfnamefont {V.}~\bibnamefont
  {Mart{\'{\i}}n-Mayor}}, \bibinfo {author} {\bibfnamefont {A.~M.}\
  \bibnamefont {Sudupe}}, \bibinfo {author} {\bibfnamefont {G.}~\bibnamefont
  {Parisi}}, \ and\ \bibinfo {author} {\bibfnamefont {J.~J.}\ \bibnamefont
  {Ruiz-Lorenzo}},\ }\href {\doibase 10.1088/0305-4470/32/1/004} {\bibfield
  {journal} {\bibinfo  {journal} {Journal of Physics A: Mathematical and
  General}\ }\textbf {\bibinfo {volume} {32}},\ \bibinfo {pages} {1} (\bibinfo
  {year} {1999})}\BibitemShut {NoStop}%
\end{thebibliography}%

\clearpage
\setcounter{figure}{0}
\makeatletter
\renewcommand{\thefigure}{S\@arabic\c@figure}
\onecolumngrid
\section*{SUPPLEMENTAL MATERIAL -- Quantum-critical scaling at the Bose-glass transition of the 3d diluted Heisenberg antiferromagnet in a field}

Here we provide supplemental data invoked in the main text as an itemized list, detailing what aspect the presented data help clarify.

\begin{enumerate}
\item
Fig.~\ref{fig:S1} shows that the finite-size $\xi/L$ for $L\leq 20$ have essentially reached saturation at the lowest accessible temperature $t=10^{-3}$ -- the data correspond to temperature scans close to the quantum-critical trajectory for $x=0.3$ (Fig.~\ref{fig:S1}(a)) and $h=3$ (Fig.~\ref{fig:S1}(b)). A similar behavior is observed throughout the phase diagram of the system. Therefore we can confidently use the critical dilutions or magnetic fields estimated at $t=10^{-3}$ to reconstruct the ground-state phase diagram, shown in Fig.~1 of the main text.

\begin{figure}[!h]
	\includegraphics[width=12cm]{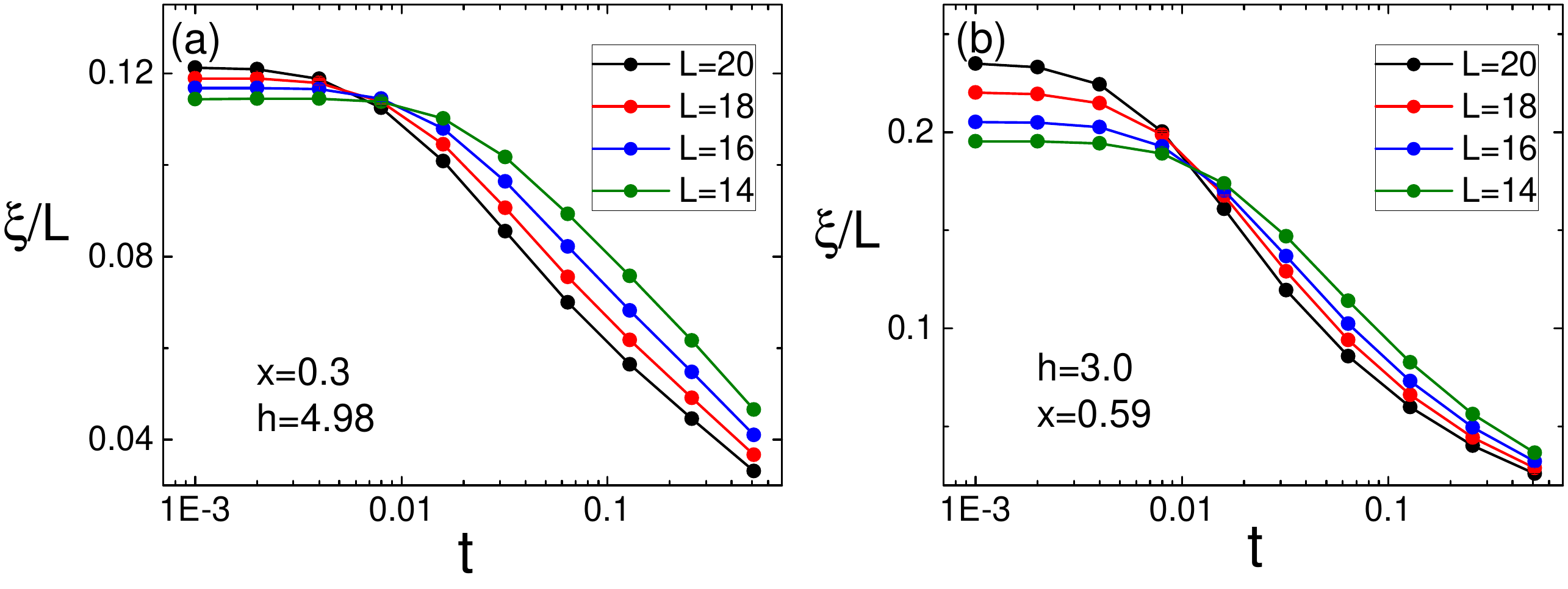} 
	\caption{Temperature dependence of the correlation length $\xi$ for finite-size systems
near the SF-BG transitions at $x = 0.3$ and $h = 4.98$ (in (a)), and at $h = 3$ and $x=0.59$ (in (b)).  }
	\label{fig:S1}
\end{figure}

\item Fig.~\ref{fig:S2} shows typical scaling plots of $\xi/L$ at $t = 10^{-3}$ -- for $x=0.3$ (Fig.~\ref{fig:S2}(a)) and $h=3$ (Fig.~\ref{fig:S2}(b)) -- allowing us to estimate the critical fields and dilutions.

\begin{figure}[!h]
	\includegraphics[width=12cm]{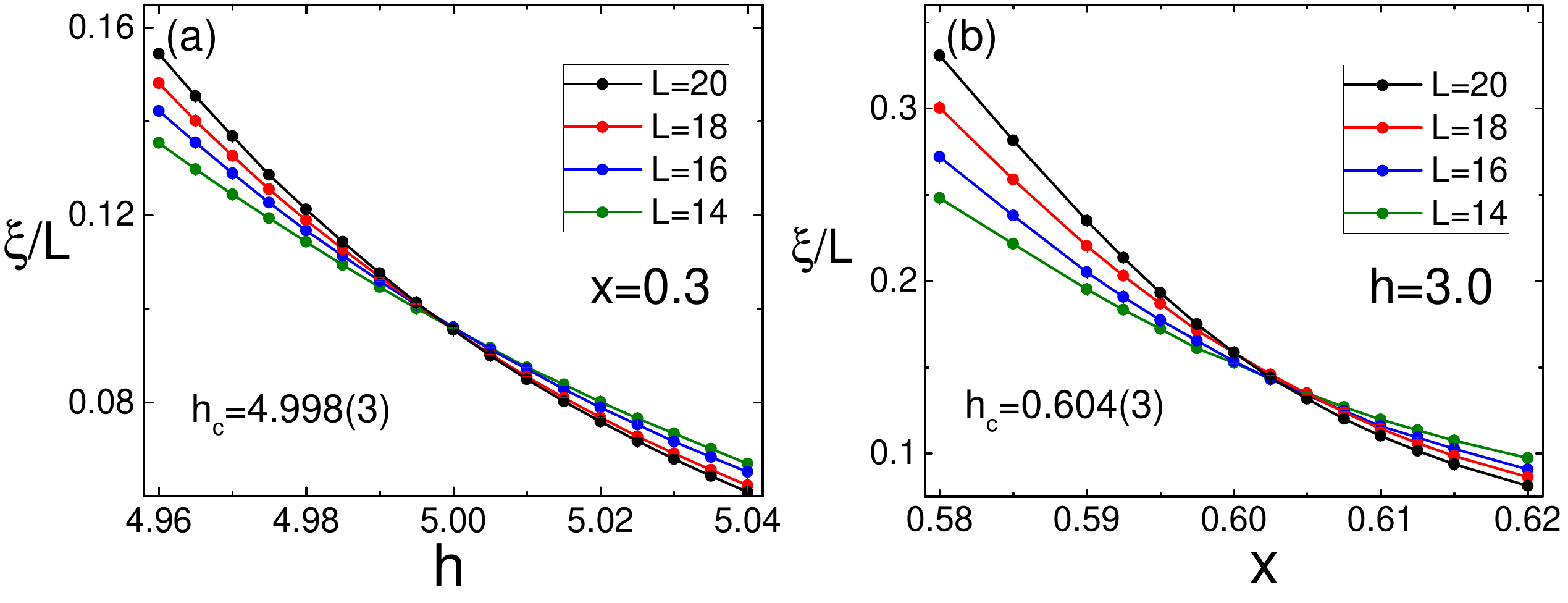} 
	\caption{Finite-size scaling of the correlation length $\xi$ for the SF-BG transitions at $t = 0.001$ by varying $h$ at $x = 0.3$ (in (a)), and by varying $x$ at $h = 3$ (in (b)). The critical points, $h_c=4.998(3)$ and $x_x=0.604(3)$, are determined by
the crossing points of $\xi/L$.
	}
	\label{fig:S2}
\end{figure}

\item {Fig.~\ref{fig:S3} shows the scaling plots of the correlation length and squared order parameter for the dilution-induced magnetic transition at zero field, using the percolation threshold as critical point $x_c(0) = x_p\approx0.6884$ and the critical exponents of 3d percolation, $\nu\approx0.876$ and $\beta\approx0.418$. Figs.~\ref{fig:S3}(a,b) show that simple collapse is not realized under this choice of critical point and critical exponents. On the other hand, if corrections to scaling are introduced, a good collapse can be obtained as shown in Figs.~\ref{fig:S3}(c,d), suggesting that the zero-field transition is indeed compatible with simple geometric percolation. To include corrections to scaling, we introduce the following general scaling Ans\"atze
\begin{equation}\tag{S1}
\label{Eq:ScalingCorrection}
\xi = L (1+c^\prime L^{-\omega}) F_{\xi}(L^{1/\nu}|x-x_c|+b^\prime L^{-\omega})~~~~~~~ m^2_s = L^{-2\beta/\nu}(1+c L^{-\omega}) F_{m^2_s}(L^{1/\nu}|x-x_c|+b L^{-\omega}).
\end{equation}
Here 
$\omega$ is related to the subleading irrelevant RG eigenvalue, and $\omega \approx 1.6(1)$ as estimated for 3d percolation \cite{Ballesterosetal1999}. $b$, $c$, $b^\prime$, and $c^\prime$ 
are fitting parameters. Note that including corrections gives more freedom for fitting, and these parameters are not uniquely determined. But for a reasonable fitting, we expect that the correction term should be subleading ones, namely, $cL^{-\omega}$, $c^\prime L^{-\omega} \lesssim 1$ and $b L^{-(\omega+1/\nu)}$, $b^\prime L^{-(\omega+1/\nu)} \lesssim \Delta x_c$, where $\Delta x_c$ denotes the shift of the critical point due to adding the correction. These criteria should apply for the biggest system sizes we considered, in order for corrections to scaling to be faithfully estimated by our finite-size data.
One can easily check that the fitting parameters in Figs.~\ref{fig:S3}(c,d) satisfy these criteria, although the finite-size correction to the scaling of $m_s^2$ is sizable.}

\begin{figure}[!h]
	\includegraphics[width=12cm]{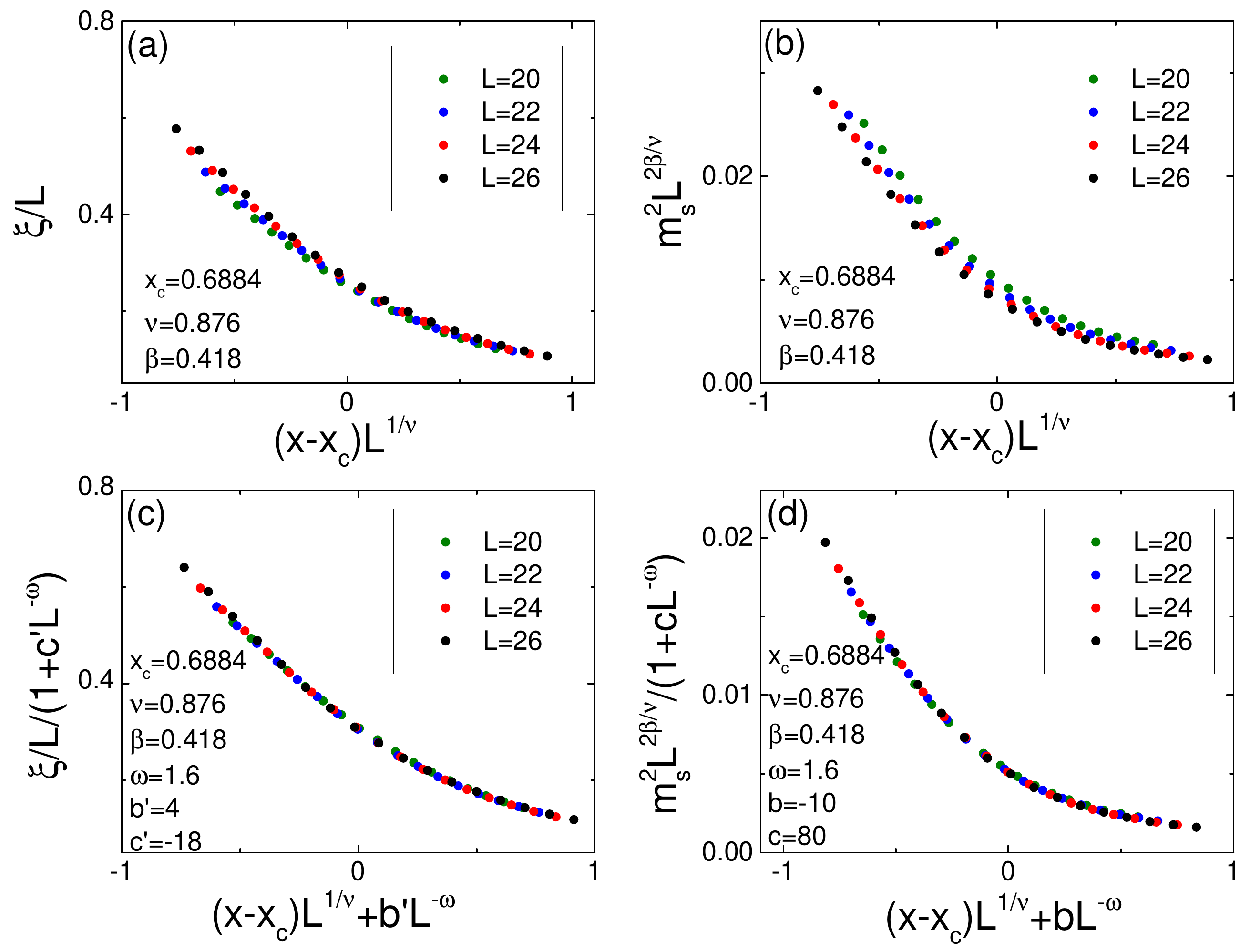} 
	\caption{Scaling plots of $\xi$ and $m_s^2$ at the zero-field induced transition with corrections, obtained after imposing that the transition occurs at $x_c=x_p$ and possesses 3d percolation exponents $\nu=\nu_p$ and $\beta=\beta_p$. We use general scaling functions, defined as in Eq.~\ref{Eq:ScalingCorrection}. Fitting coefficients are shown in the figures. (a,b) Scaling plots without corrections to scaling. (c,d) Same scaling plots as in (a,b), but including corrections to scaling.}
	\label{fig:S3}
\end{figure}

\item Fig.~\ref{fig:S4} shows the plots of $h_c(t)-h_c(0)$ for $x=0.3$ (Fig.~\ref{fig:S4}(a)) and $x_c(t)-x_c(0)$ for $h=3$  (Fig.~\ref{fig:S4}(b)) vs. t in log-log scale. These plots highlight that the asymptotic power-law behavior appears only in the very-low-$t$ range ($t \lesssim 10^{-2}$) .
\begin{figure}[!h]
	\includegraphics[width=12cm]{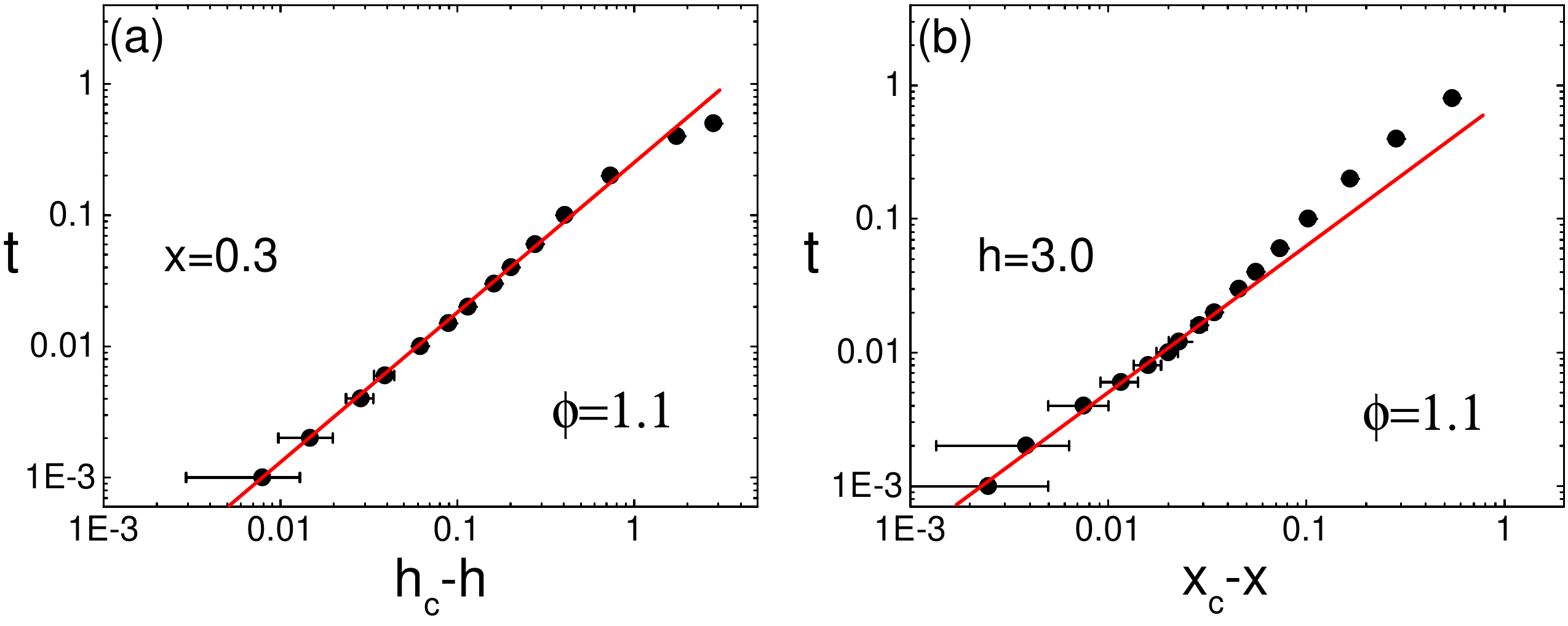} 
	\caption{Scaling of the finite-temperature critical points log-log scale for $x = 0.3$ (a),  and for  $h = 3$ (in (b)). Lines are fit to $g_c(t) = g_c(0) - A t^{1/\phi}$ with (a) $g=h$  (a), and (b) $g=x$. The $g_c(0)$ values have been extracted from a windowing analysis, see Fig.~\ref{fig:S5}. }
	\label{fig:S4}
\end{figure}

\item Fig.~\ref{fig:S5} show a windowing analysis of the fit of the finite-temperature critical points to the function $g_c(t) = g_c(0) - A t^{1/\phi}$, with fitting parameters  $g_c(0), A$ and $\phi$. Data are fitted within a temperature window $t\in [10^{-3},w]$ with variable $w$. In the case $x=0.3$ we clearly observe that the fitting parameters $h_c$ and $\phi$ stabilize upon decreasing $w$ to what appears to be their asymptotic values. The case $h=3$ is instead more delicate, as the fitting coefficient appear to still slightly drift. We take nonetheless the values for the next-to-smallest $w$'s as our best estimates; Fig.~3 of the main text shows nonetheless that the quality of the fit is excellent by using these values.

\begin{figure}[!t]
	\includegraphics[width=12cm]{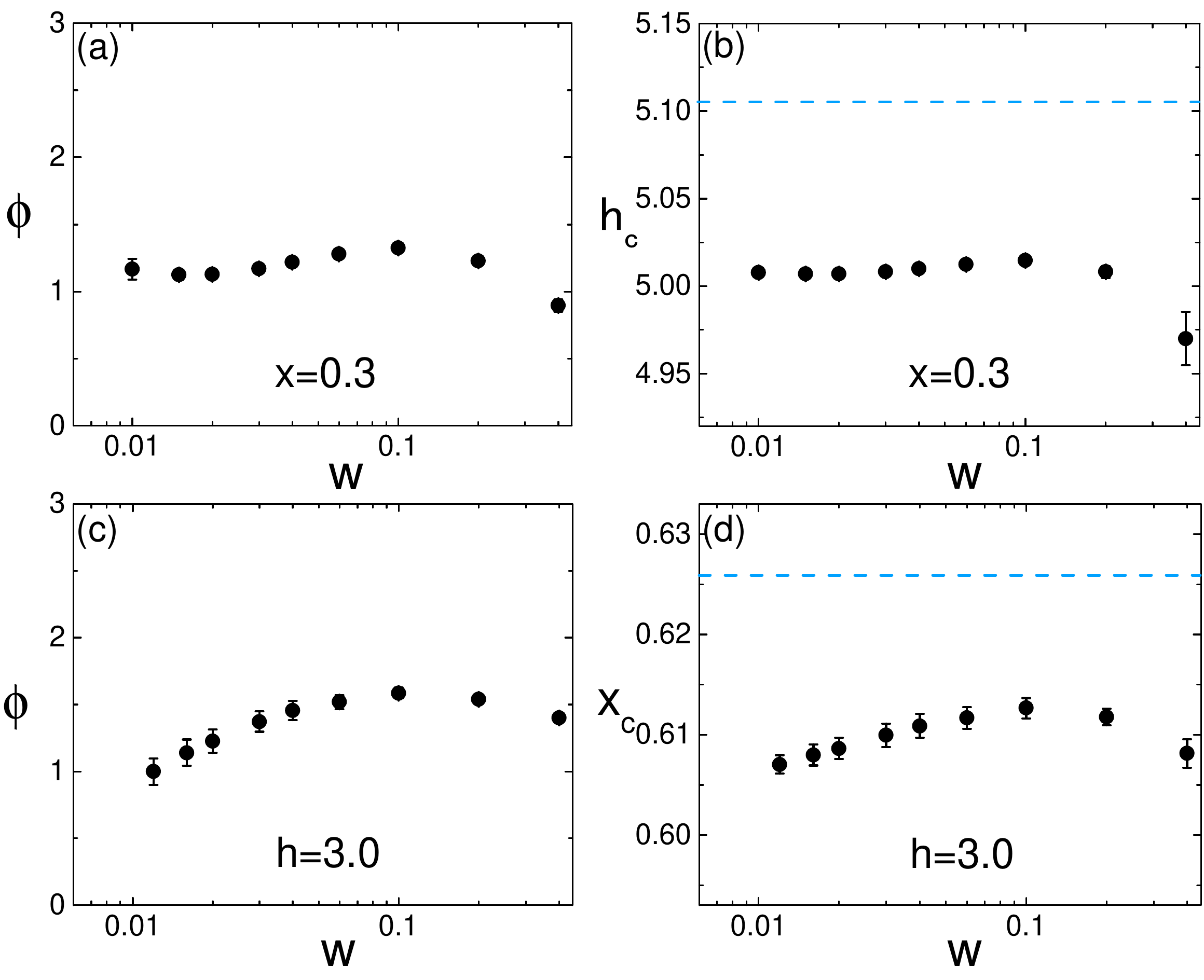} 
	\caption{Windowing analysis of the finite-temperature transition lines,  $h_c(t)$ at $x = 0.3$ (in (a),(b)), and $x_c(t)$ at $h = 3$ (in (c),(d)). We show the dependence of the extracted $\phi$ and $g_c(0)$ values by fitting the data to $g_c(t) = g_c(0) + A t^{1/\phi}$ within the temperature window $t\in[10^{-3},w]$. The blue dashed lines in (b) and (d) mark the $g_c(0)$ values obtained by the same fit (to intermediate-temperature data), in which we fix $\phi$ to the value $\bar{\phi}=2$ (see main text).}
	\label{fig:S5}
\end{figure}

\item Fig.~\ref{fig:S6} shows the scaling plot for $m_s^2$ at the SF-BG transition for $x=0.3$. As shown in Fig.~\ref{fig:S6}(a), without including corrections to scaling, the best scaling collapse is obtained using exponents $\nu = 0.90(5)$ and $\beta = 1.20(9)$; the latter is clearly incompatible with the $\beta$ exponent of 3d percolation, $\beta_p = 0.418(1)$. To correctly contrast the SF-BG transition with the 3d percolation one, we tried to impose the critical exponents of percolation $\nu_p$ and $\beta_p$, including at the same time corrections to scaling along the lines of  Eq.~(\ref{Eq:ScalingCorrection}), and with the percolation exponent $\omega = 1.6$ for the correction terms. As shown in Fig.~\ref{fig:S6}(b), one can achieve a reasonable data collapse by appropriately adjusting the $b$ and $c$ parameters; yet the $c$ parameter is anomalously large,  so that $c L^{-\omega}\gg 1$ even for the biggest system sizes we considered. According to the argument made above, this makes the scaling analysis with scaling corrections unreliable, indicating that $\beta$ exponent of the SF-BG transition must deviate largely from the 3d percolation one.

\begin{figure}[!h]
	\includegraphics[width=12cm] {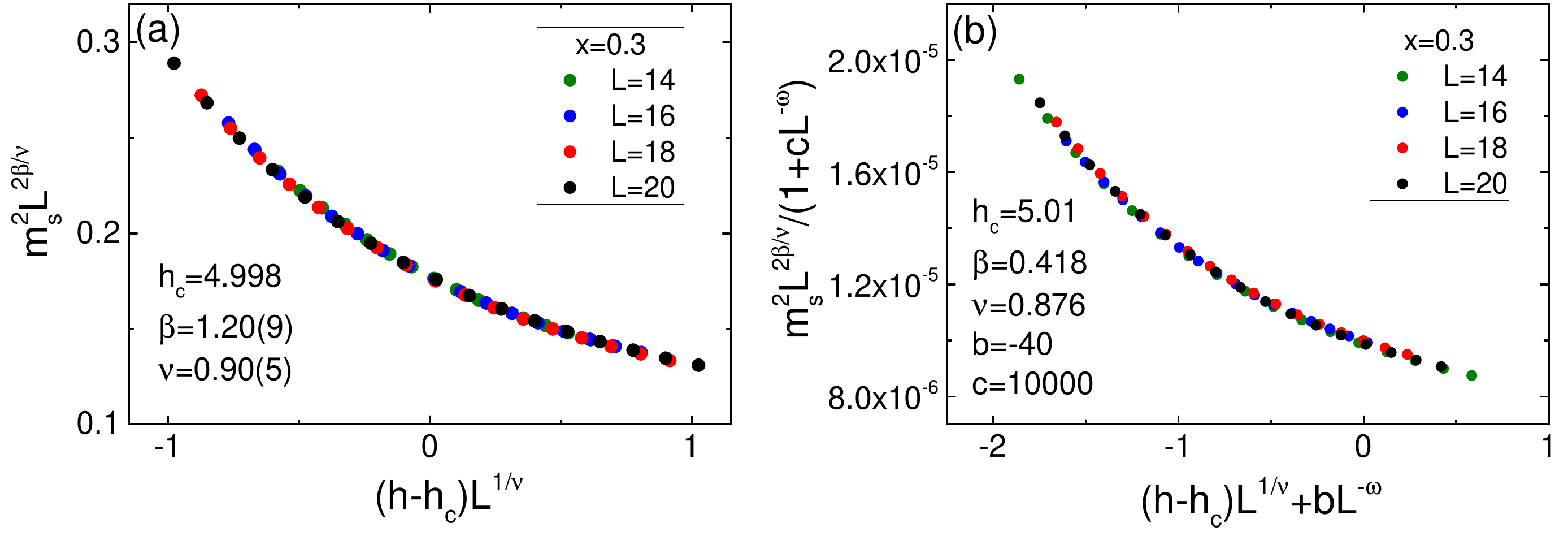} 
	\caption{Scaling plot for $m_s^2$ at the SF-BG transition for $x=0.3$. (a) Scaling plots without corrections to scaling. (b) Same scaling plots as in (a), but including corrections to scaling.}
	\label{fig:S6}
\end{figure}

\end{enumerate}

\end{document}